# Effect of reaction temperature on nascent carbonaceous particles from toluene shock-tube pyrolysis: Insights from FTIR and Raman spectroscopy

Meysam K. Rezaeian[a], Can Shao[a*], Jürgen Herzler[a], Mustapha Fikri[a], Greg J. Smallwood[b], Christof Schulz[a]

[a]*EMPI, Institute for Energy and Materials Processes – Reactive Fluids and CENIDE, Center for Nanointegration Duisburg-Essen, University of Duisburg-Essen, 47048 Duisburg, Germany*
[b]*Metrology Research Centre, National Research Council Canada, Ottawa, ON K1A 0R6, Canada*

**Abstract**

The transition from gaseous precursors to nascent solid particles and their subsequent structural maturation were investigated in single-pulse shock-tube experiments using *ex situ* Fourier-transform infrared (FTIR) and Raman spectroscopy of sampled products. A mixture of 2 % toluene in argon was pyrolyzed at 2.0 ± 0.1 bar with temperature plateau times of 2.0 ± 0.2 ms over the 1450–1800 K reaction temperature range. *In situ* laser extinction measurements indicate the onset of particle formation at 1570 K. At this temperature, Raman spectra exhibit emerging D and G bands, and transmission electron microscopy (TEM) reveals the disappearance of poorly-defined structures, identifying 1570 K as the phase-transition reaction temperature. Approaching this reaction temperature, Raman spectra show a rapid disappearance of sp-hybridized C ≡ C bonds. At 1670 K reaction temperature, a maximum in primary particle diameter and a decrease in structural disorder inferred from Raman spectroscopy are observed, defining the ordering threshold. Deconvolution of the FTIR spectra enables separation of in-ring C=C stretching vibrations from isolated and ring-conjugated side-chain C=C modes. The in-ring C=C band is used to normalize aliphatic and aromatic C–H vibrations. FTIR analysis reveals ring-edge structures associated with electron-localization - sites, including bay regions, five-membered ring defects, and benzylic positions, indicating a radical-rich environment below the phase-transition temperature. Between the phase-transition and ordering-threshold temperatures, K-regions and armchair structures associated with electron delocalization and thermal stability increase. The emergence of these electronic and structural characteristics highlights the critical role of radicals in soot inception and early structural ordering.

*Corresponding author.

## Introduction

Carbonaceous nanoparticles generated by incomplete combustion of fossil fuels pose challenges for human health as well as the climate [1, 2]. Soot formation has been extensively studied over the last decades, however, the initial transition from gas-phase precursors to the first particles, referred to as soot inception, is still not fully understood [3]. Pioneering research has indicated that during inception, highly reactive liquid-like particles are formed [4].

Several hypotheses for the soot inception mechanism have been proposed and studied both experimentally and theoretically [3, 5]. One hypothesis of soot inception centers on non-covalent moderate-size (molecular masses ~400–800 u) polycyclic aromatic hydrocarbons (PAHs) clustering together by Van der Waals forces [6-8], and modeling emphasizes the reversibility of the dimerization step that leads to further growth [9, 10]. A second pathway is chemical inception, where resonance-stabilized hydrocarbon radical-chain reactions are involved [11]. Aerosol mass spectrometry experiments together with high-level quantum chemistry calculations have shown evidence of this pathway [12]. Work by Adamson et al. [13] using high-resolution mass spectrometry detected aliphatically-bridged multi-core PAHs that demonstrated radical-mediated cross-linking. Recent STM/AFM measurements coupled with quantum molecular dynamics simulation [14] indicated that a triplet π-diradical might be the key of soot inception. Chain reactions between these precursors produces bonds that are strong enough to stabilize PAH dimers at flame temperatures. Edge-localized radicals at zig-zag sites further enhance intermolecular attraction, a concept long advocated and supported by electronic-structure analyses that treat those edges as sites bearing localized unpaired π electrons (radical centers) with pronounced diradical character [14-16].

Fourier transform infrared spectroscopy (FTIR) produces spectra that serve as fingerprints that incorporate information regarding the functional groups, bonds, and structures[17], and has been used for characterization of soot particles [18, 19]. In FTIR, the concurrent presence of multiple peaks at 3000–3100 $cm^{-1}$ (aromatic C–H stretch) and 1580–1600 $cm^{-1}$ (C=C stretch) confirms the presence of aromatics. The spectral range ~900–700 $cm^{-1}$ covers out-of-plane (oop) aromatic C–H bending modes, i.e., solos (isolated hydrogen atoms), duos (two adjacent hydrogen atoms), trios (three adjacent hydrogen atoms), and quartets (four adjacent hydrogen atoms) around aromatic rings [20-24]. Cain et al. provided evidence for the presence of aliphatic hydrocarbons on the surface of nascent soot using micro-FTIR [18]. They also observed that the quantity of aliphatic hydrocarbons increased with increasing flame temperature. Later, they hypothesized that this observation is linked to persistent free radicals that serve as active sites on the aromatic surface for aliphatic species to bind [19].

Raman spectroscopy is widely use to characterize carbonaceous particles by measuring the degree of structural order [25, 26] or the distance between the defects in the $sp^2$ network as a function of flame temperature [27]. Additionally, studies have been conducted to connect the presence of peaks in Raman spectra at 1900–2200 $cm^{-1}$ to other structures such as sp bonds like polyynes (–C≡C–C≡C–), and sp2 bonds in cumulenes (=C=C=C=C=) [28-32]. Le et al. [33] applied inline Raman spectroscopy for soot particles in an aerosol flow sampled from a low-pressure pre-mixed ethylene flame, which revealed a large abundance of sp carbon chains in disordered soot. The potential role of polyynes and cumulenes in the inception stage was underscored after finding a large portion of carbon content in incipient soot particles are sp carbon chains that link graphitic domains of soot precursors [34].

Deconvolution of FTIR measurements to unveil the composition of soot has been employed in only a few studies. Russo et al. deconvolved the C–H stretch and aromatic oop signals in FTIR measurements of hydrogen-rich and hydrogen-poor carbonaceous material (overall H/C = 0.06−0.95), including aromatic precursors, flame-generated samples, and commercial carbon black. They quantified the H/C ratio based on FTIR signal peak heights [35]. Basta et al. [36]deconvolved FTIR spectra obtained from atomic hydrogen-exposed flame-synthesized carbon nanoparticle films (young vs. mature soot from premixed ethylene/air flames). They separated aromatic and specific aliphatic sub-bands within the C–H-stretch region, which enabled them to track the amount of aliphatic C–H (rise–peak–drop) and aromatic C=C (continuous decrease) under hydrogen exposure.

Transmission electron microscopy (TEM) is one of the pillars of nanomaterials diagnostics. It has been used to confirm or strengthen the findings in many soot studies. Aberration-corrected TEM images resolve graphene fringe length, interlayer spacing, and defect topology, letting researchers watch 1-nm aromatic "islands" merged into loosely stacked graphene layers during engine cycles [37] and even reconstruct full 3D particle architecture employing atomic force microscopy (AFM) [4].

Most studies on soot inception have been conducted on flame-generated soot, where the inception zone is typically narrow and affected by steep temperature gradients, making measurements in the transition zone from molecular species to incipient particles difficult. Alternatively, shock-tube experiments provide the opportunity to resolve the relevant reaction steps not on a spatial but on a temporal axis taking advantage of the high level of control of temperature, pressure, and equivalence ratio, including oxygen-free pyrolytic conditions. A previous study [38] identified an inception only condition, where the minimum temperature for soot particle detected by time-resolved LII was 1524 K, 3.2 ms after the arrival of the reflected shock wave, during pyrolysis of 1 % toluene in argon in a shock tube. Due to the low reaction temperature and short reaction times, particles formed under these conditions retain their initial morphology without aging, as supported by the *ex situ* analysis. However, at higher reaction temperatures, soot inception is very fast followed by rapid growth and carbonization. Single-pulse shock tubes (SPST) [39] can be employed to study soot formation under variable well-controlled temperature and pressure conditions, providing an environment for analyzing soot inception with optical [38] and sampling techniques. In a shock tube operated in single-pulse mode, the rapid cooling after exposing the reaction for a limited time of a few milliseconds to a "plateau temperature", preserves the reactive mixture from reheating through secondary waves.

In this work, carbonaceous particles formed in a single-pulse shock tube during toluene pyrolysis at different temperatures were investigated by FTIR and Raman spectrometry, and with additional nanostructure information by TEM imaging. By coupling these *ex situ* diagnostics, a phase-limiting temperature (formation of the first solid particles), in addition to an ordering threshold temperature (at which edge defects and surface disorder drop), are identified, and the impact of temperature on the functional groups and edge structures are studied. The temperature-resolved characteristics of structures in terms of hosting free electrons are discussed and utilized to map the probability of electron localization/delocalization at temperatures in which inception and various degrees of maturation are observed.

## 1. Experimental setup and analysis methodology

### *1.1 Single-pulse shock-tube facility*

Pyrolysis experiments were carried out in a SPST facility at the University of Duisburg-Essen, described in detail previously [39]. In summary, this facility is operated to create sudden heating of the reactive gas mixture by the incident and reflected shock waves while avoiding reheating by secondary waves reflected from the other end of the shock tube. For the single-pulse mode, the shock tube is connected to a dump tank (volume: 0.35 $m^3$) via a ball valve. The dump tank is filled with nitrogen at the same pressure as the driven section of the shock tube, and the valve is opened just before an experiment is started. The driver section (length: 2.65 m) and the driven section (6.32 m) of the 80-mm diameter shock tube are separated by an aluminum diaphragm with thicknesses between 60 and 80 μm. The incident shock-wave velocity is determined by measuring the arrival of the incident shock wave at five pressure transducers located at various distances from the end wall. The final pressure transducer is located at 7.5 mm from the end wall, at the same location where a HeNe laser passes through the test section for light extinction measurements.

Based on the incident shock velocity and the initial pressure ($p_1$), the temperature and pressure ($T_5$ and $p_5$) just behind the reflected shock are calculated based on the Chemkin-Pro shock-tube model [40]. To obtain the corresponding temporal temperature profiles, the pressure profile measured at the final pressure transducer and the calculated $T_5$ are used as inputs to a 0D constant-pressure batch reactor simulation in transient mode in Chemkin-Pro coupled with the CRECK mechanism [41]. The calculated $T_5$ has an uncertainty of ±15 K [42], which is equivalent to ~1 % of the lowest temperature (around 1496 K) investigated in this study.

A selection of temperature profiles within the temperature range under study is shown in Fig. 1. The temperature profiles with $T_5$ higher than 1670 K begin with a sharp drop due to the endothermicity of pyrolysis and then reach a "plateau temperature," constant (within ±30 K) for ~2 ms (attributed to further reactions and non-ideal gas dynamics interactions). These plateau temperatures are averaged over the 2 ms interval and used to label the experiments as the relevant reaction temperatures. Fast cooling then starts after 1.8–2.2 ms.

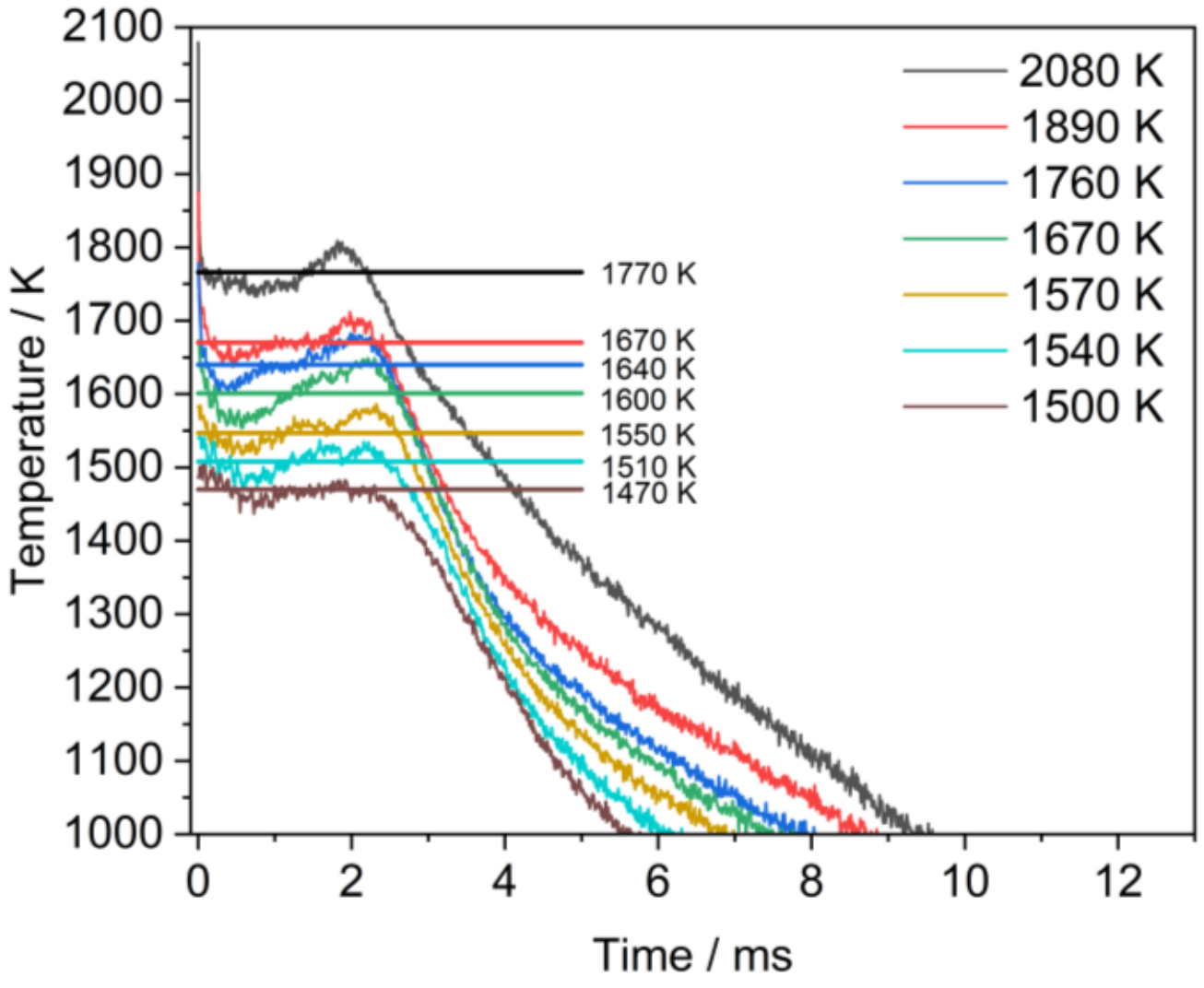

Fig. 1. Temperature profiles labelled by $T_5$ rounded to the nearest 10 K in the legend and plateau temperatures (horizontal lines) after arrival of the reflected shock at the laser location.

Uncertainty analysis for similar pressure-derived temperature profiles in a rapid compression machine showed that the relative uncertainty of the compressed temperature (no reaction) does not exceed 0.7 % [43]. In prior work, two-line CO absorption temperature measurements in the present facility found a good agreement between the simulated and measured temperature profiles [44]. On this basis, and as a conservative choice, we apply a factor of three and adopt error bars of ±3 % on the reported temperature profiles and subsequently for plateau temperatures used for labeling the experiments (Figure S1 in the supplementary material). For all experiments, a constant pressure of 2.0 ± 0.1 bar is achieved and the plateau temperature (averaged temperature over plateau time) was kept for 2.0 ± 0.2 ms followed by a cooling phase (cooling rate: 100–200 K/ms) to ambient temperature.

To monitor particle formation behind the reflected shock waves, light extinction was measured as a function of time with a HeNe laser (Thorlabs, HNL225RB) at 633 nm. The signal was detected by a Si photodiode (Thorlabs, SM05PD1A) with a 630 ± 10 nm interference filter, which was mounted onto a 50-mm diameter integrating sphere (Thorlabs, model IS236A-4). The measured extinction was converted to normalized optical densities $D_{633nm}$ as described by Drakon et al. [42]:

$$D_{633nm} = \ln(I_0/I) \, / \, l[C]_0 \qquad (1)$$

where $I_0$ and $I$ are the incident and transmitted light intensity, $l$ is the absorption path length in m, and $[C]_0$ is the initial total carbon concentration in the reactive mixture, in mol/m$^3$. The normalization is performed to provide comparability among the previous optical density measurements using different mixtures and shock-tube diameters as the path length.

All experiments were conducted using 2 % toluene in argon at 2.0 ± 0.1 bar over a plateau temperature range of 1450–1800 K. Toluene was chosen for consistency with our prior work [38, 45]. Additionally, soot formation during toluene pyrolysis is sufficiently rapid and causes strong attenuation of the laser signal during the plateau period, even at relatively low temperatures, enabling a systematic investigation of early inception.

### *1.2 Ex-situ sample characterization*

After each experiment, carbonaceous particles were collected on a glass fiber filter (pore size 0.7 µm, Whatman GF/F) mounted in an exhaust pipe, by pumping the exhaust aerosol from the shock tube for 10 min. The particles are then prepared for further characterization as described below.

Deposited materials from the filters are immersed in ethanol (⩾99.8 % purity) and ultrasonicated for 10 min. Samples from the resulting suspension are deposited onto TEM grids and dried, followed by morphological analysis of the images acquired with TEM (Cs-corrected JEOL JEM-2200FS, 200 kV acceleration voltage). Primary particle diameter distributions are obtained from TEM images using a custom Python-based image analysis workflow built upon the SimpliPyTEM library [46]. Individual primary particles are manually selected, and their projected outlines are quantified by ellipse fitting, accounting for non-spherical particle projections. The particle diameter is defined as the arithmetic mean of the fitted major and minor ellipse axes.

Raman spectra are acquired at room temperature under ambient atmosphere with a Renishaw inVia micro-Raman spectrometer at wavenumbers between 300 and 3200 cm$^{-1}$, operated at 532 nm with a diffraction grating of 1800 l/mm. During the measurements the laser is focused onto the sample through the built-in microscope, producing a 10 µm spot. At low reaction temperatures, where liquid-like material is deposited on the quartz filters, measurements are carried out directly on the sample coating the filter. For higher reaction temperatures with powder-like product, the samples are transferred from the filter to a glass slide and subsequently measured. Spectra are collected at multiple points across each filter/slide to check sample homogeneity, with all measurements performed under identical optical alignment and exposure time settings. The Raman spectra measurements from a clean filter and a blank slide showed no interferences with the spectrum of the samples. Figure S2 shows the Raman spectra obtained from a clean filter.

FTIR analyses are conducted utilizing a Bruker Vertex 80 spectrometer, with standard optical components consisting of a KBr beam splitter, a DigiTect DLaTGS detector, and a 400–4000 cm$^{-1}$ light source. A resolution of 1 cm$^{-1}$ is used in all the analyses. A scan-number study (varying between 32 and 500 scans) shows scan-number independent results, thus a scan number of 32 is adopted. KBr pellets with a 13-mm diameter are created by pressing a homogenized powder of 150 mg KBr powder and 0.4 mg of the high-temperature solid sample particles (collected from the filter) with a force of 85 kN for five minutes. For the low-temperature cases, the liquid-like samples are transferred from the filter onto a clean KBr pellet. The pellet plus liquid-like sample was crushed and re-pressed to create a measurement pellet. All pellets are created and analyzed within 30 min after sample collection to minimize oxidation and aging. Measurements of pure KBr pellets are used for background correction. During the measurements, the test chamber is continuously purged with dry nitrogen to eliminate interference effects due to atmospheric $CO_2$ and $H_2O$ vapor. To evaluate whether residual volatile biases the FTIR spectra of the liquid-like soot samples, two representative samples were subjected to (i) vacuum exposure at room temperature for 1 h or to (ii) heating at 200 °C [47] for 1 h in argon.

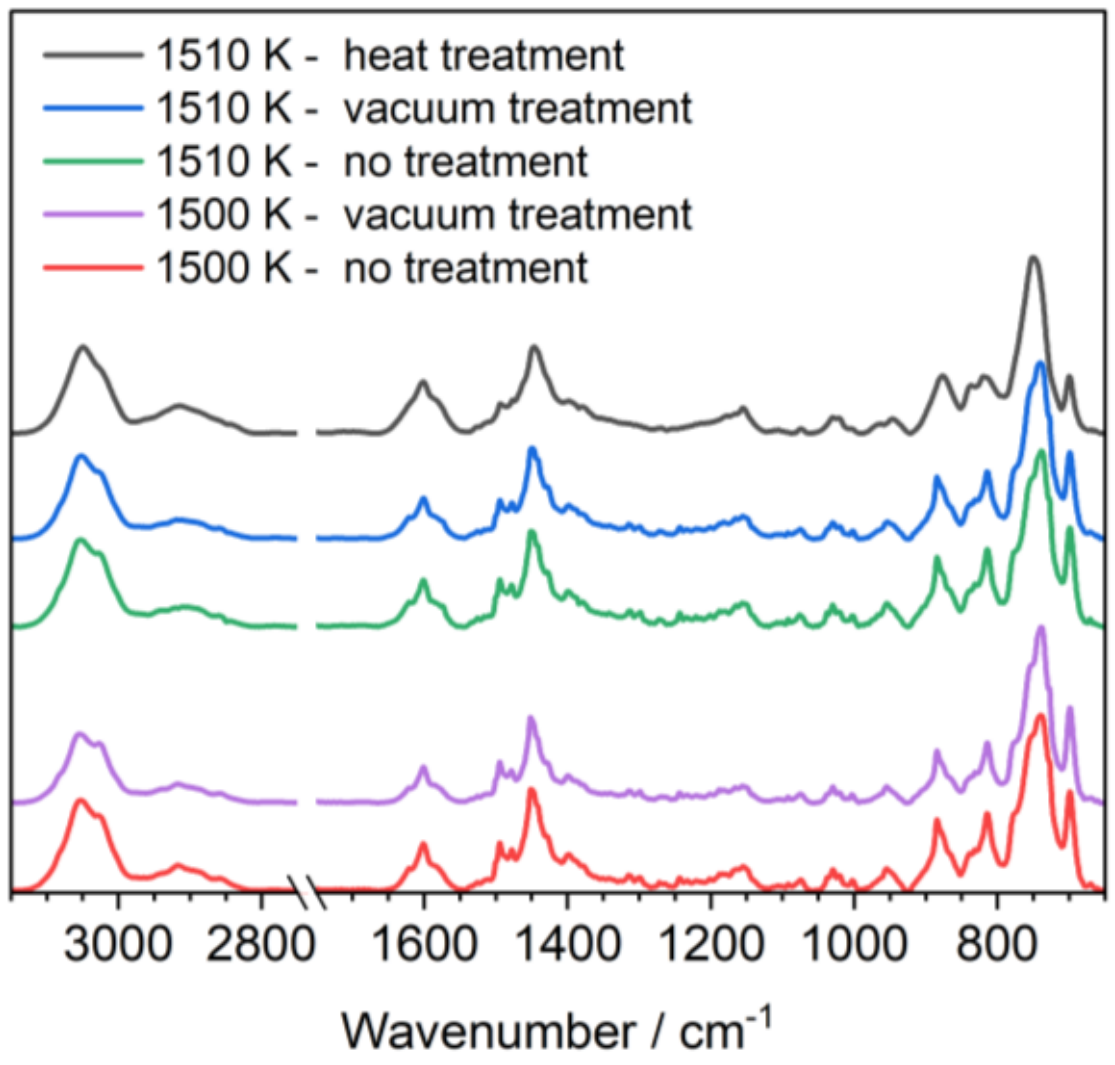

Fig. 2: FTIR spectra for two liquid-like samples (1500 and 1510 K cases) without treatment, with vacuum treatment, and with heat treatment (200 °C, argon flush; 1510 K only).

Figure 2 shows that neither of the post-treatment methods resulted in the disappearance or addition of peaks. Room-temperature vacuum treatment produced a 35 % mass loss, and a minor broadband intensity decrease was observed in the 1500 K case. This indicates that the light volatilized material in room-temperature vacuum treatment, despite having a considerable portion of the sample weight, creates no interference in FTIR measurements. In contrast, the 200 °C heat treatment produced a 40 % mass loss and caused a decrease of the aromatic C–H out-of-plane "duo" feature (by ~50 %) together with a slight increase in aliphatic C–H stretching intensity. These changes are the result of a stronger evaporative loss of condensed phase relative to vacuum treatment. The similarity of fresh and treated samples subjected to vacuum treatment at room temperature indicates that the sampling procedure, which includes the collection of the material by pumping the samples through a vacuum line for ten minutes, already removes the present (if any) residual volatile with interferences to FTIR measurements . Therefore, all FTIR spectra reported in this work were acquired on as-collected samples without additional post-treatment. During the spectral post-processing, the background subtraction and deconvolution are conducted using SpectroChemPy [48] and described along with a sensitivity analysis in detail in the supplementary material.

It should be noted that, due to experimental constraints, each measurement method has its own measurement campaign. Consistent experimental procedures were used across all campaigns to maintain comparability.

## 2. Results and discussion

This study includes a range of temperatures that produced samples from liquid-like brown material at lower reaction temperatures to powder-like black material at higher temperatures. Figure 3 shows the temporal evolution of the normalized optical densities derived from time-resolved *in situ* extinction measurements, revealing two pivotal temperatures. First at ~1570 K, we refer to as phase-limiting temperature, where the normalized optical density starts to deviate from zero, and the second at ~1640 K, where extinction suddenly increases, demonstrating significantly accelerated particle formation and growth.

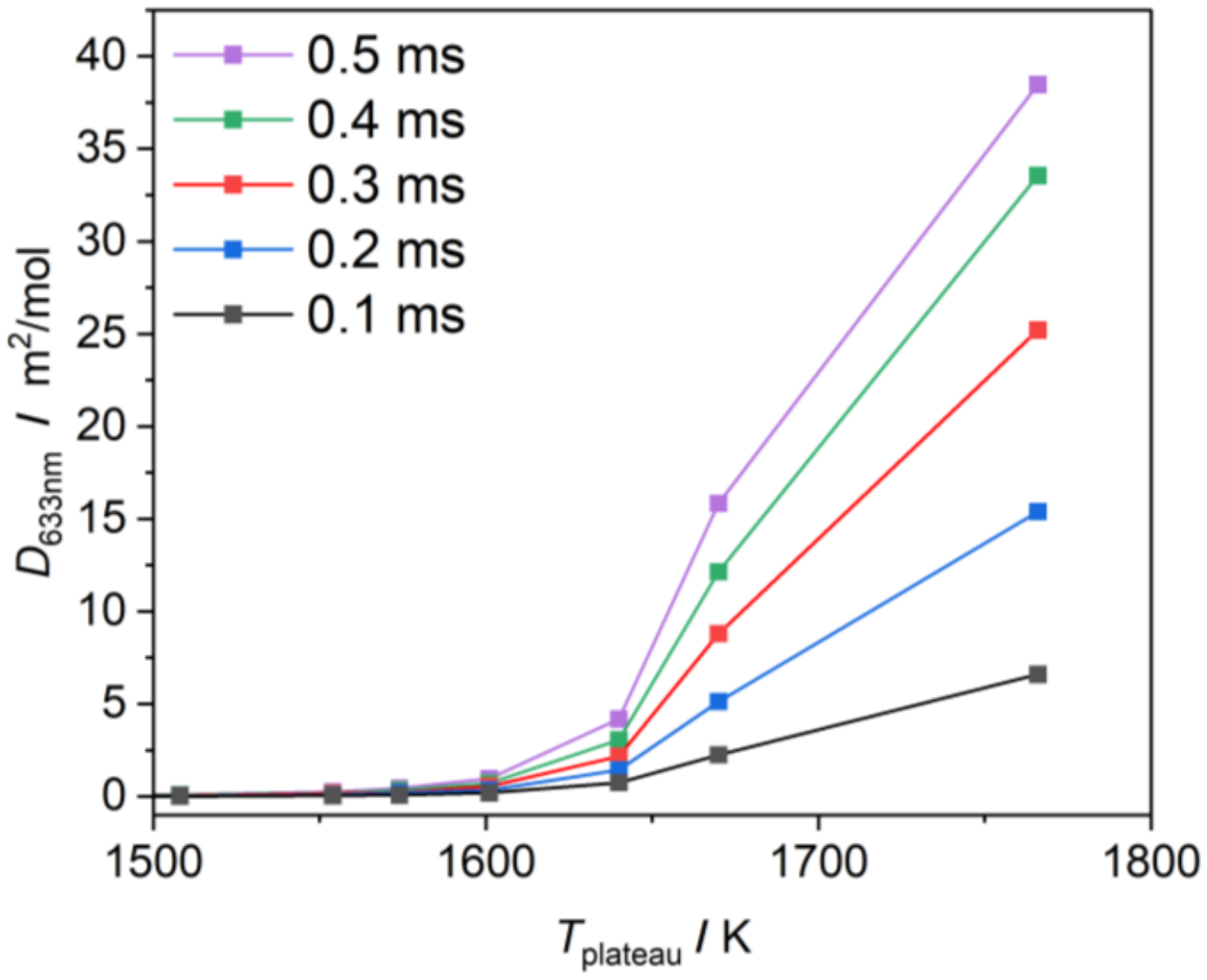

Fig. 3: Measured normalized optical densities as a function of plateau temperature for selected reaction times.

TEM measurements support the systematic change of the materials at the phase-limiting reaction temperature of 1570 K, by showing poorly-defined particle shapes at or below that temperature, along with some aggregates of spherical primary particles (seen in Fig. 4 a–f). At higher temperatures, only aggregates of spherical primary particles (Fig. 4 g–i) are observable.

The median values of the primary particle diameter ($d_p$) distribution derived from TEM images (Figure S6) are shown in Fig. 5 as a function of the plateau temperature. The primary particles formed at low temperatures (≤ 1570 K) are smaller, some with poorly-defined shapes and some for which the individual primary particles may be identified within aggregates. Conversely, above the phase-limiting reaction temperature (1570 K), they only form aggregates while maintaining the individual particle structure. Above the phase-limiting temperature, they quickly collapse into larger structures near 1670 K, followed by size reduction at higher temperatures.

Figure 6 shows Raman spectra of the materials sampled from the shock tube. For comparison with well-structured carbonaceous material formed at high reaction temperatures, we also show Raman spectra of few-layer graphene (FLG) produced in a plasma reactor [49]. The spectra are peak normalized in each sub-graph to aid in distinguishing regions with no signals from regions with small signals. Table 1 represents the corresponding peaks assigned to Raman features. In Fig. 6, one may observe that the spectra recorded start to show D (~1350 $cm^{-1}$) and G (~1580 $cm^{-1}$) bands for cases with temperatures at and above ~1570 K. This observation indicates the formation of the first $sp^2$ structures in sufficient amount to be detected, justifying ~1570 K as the phase-limiting temperature. The first weak signs of D and G bands can be found at 1510 K. Interestingly, the corresponding $T_5$ (~1540 K) value is marked as the limit at which the first time-resolved laser-induced incandescence (LII) signals have been detected during toluene pyrolysis behind the reflected shocks [38]. On the other hand, a series of peaks are observed at 1800–2500 $cm^{-1}$ in the Raman spectra around the phase-limiting temperature, which correspond to the bands attributed to polyynes ($C_1$, 1800–1940 $cm^{-1}$), and polycumulenes ($C_2$, 1980–2100 $cm^{-1}$) [28-30] in amorphous carbon [32]. These groups are introduced as bridges to the neighboring $sp^2$ cluster/PAH layers in nascent soot [32-34]. The D, G, and 2D (at 2700 $cm^{-1}$) are observed simultaneously only for the FLG case, indicating the presence of highly ordered structures of graphene layers.

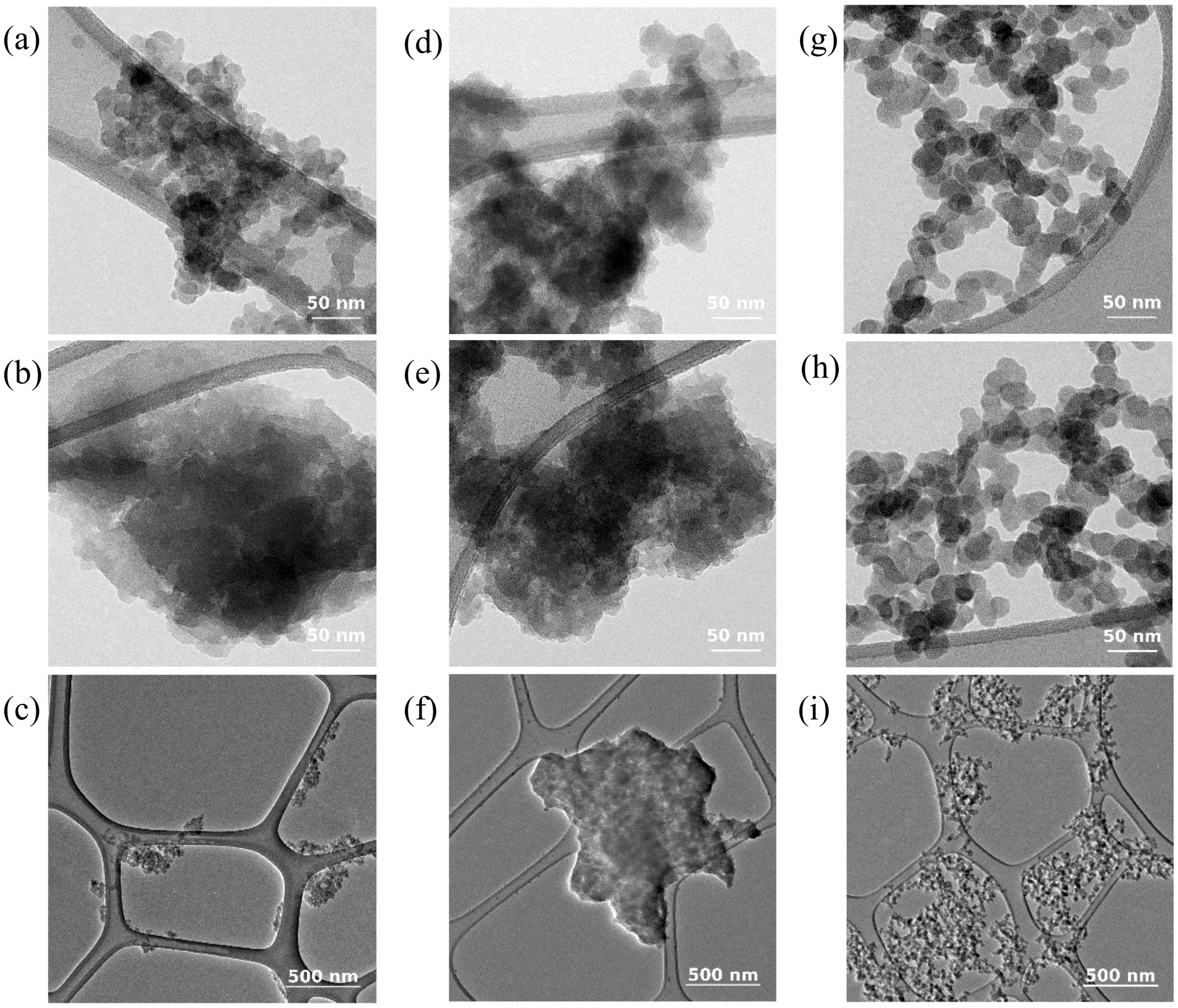

Fig.4: Representative TEM images from sampled particles at two magnification levels at $T_{plateau}$ = 1550 K (a–c), 1570 K (d-f), as cases with reaction temperatures at or below the phase-limiting temperature, and 1670 K (g–i) as a case well above the phase-limiting temperature.

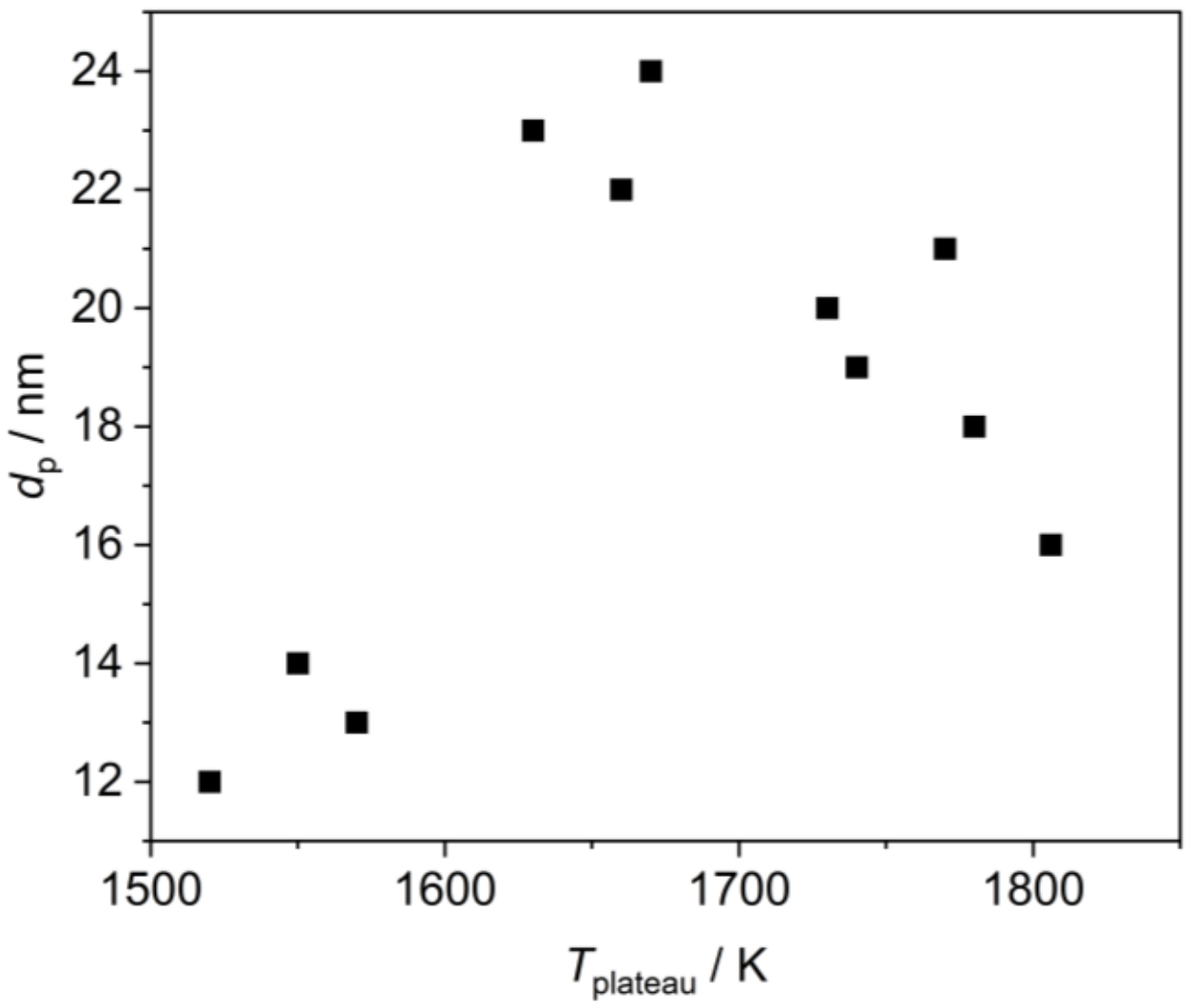

Fig. 5: Variation of the median of the primary particle diameter ($d_p$) distribution with plateau temperature.

Table 1: Raman peak assignment.

| Peak position / cm⁻¹ | Assignment |
|---|---|
| 1360 (1350–1370) | $D_1$: Activated modes from symmetry-breaking defects in a graphitic lattice, representative of edge defects [26, 33]. |
| 1620 (1599–1624) | $D_2$: Lattice vibrations from structurally perturbed misoriented surface/stacking features or structural disorder at the surface of a graphitic crystal [25, 26]. |
| 1500 (1489–1545) | $D_3$: Band originating from the amorphous carbon fraction of soot [26, 33]. |
| 1200 (1127–1208) | $D_4$: Shoulder on $D_1$, tentatively attributed to $sp^2$–$sp^3$ bonds or C–C and C=C stretch of polyene-like structures, or mixed $sp^2$–$sp^3$ C–C stretch [26, 33]. |
| 1580 (1571–1598) | G: G mode of the graphitic lattice (aromatic $sp^2$ in-plane stretch) [26, 33]. |
| 1800–1940 | $C_1$ (A mode): Fingerprint band of sp carbon and $sp^2$ polycumulene chains stretch [28, 33, 34]. |
| 2000–2100 | $C_2$: In-phase stretch of polyynes along sp chains [28, 34]. |
| ~2220 | (B-mode): Alternating (out-of-phase) stretch of polyynes neighboring the A mode [28]. |
| ~2380 | Higher mode of polyyines [28]. |

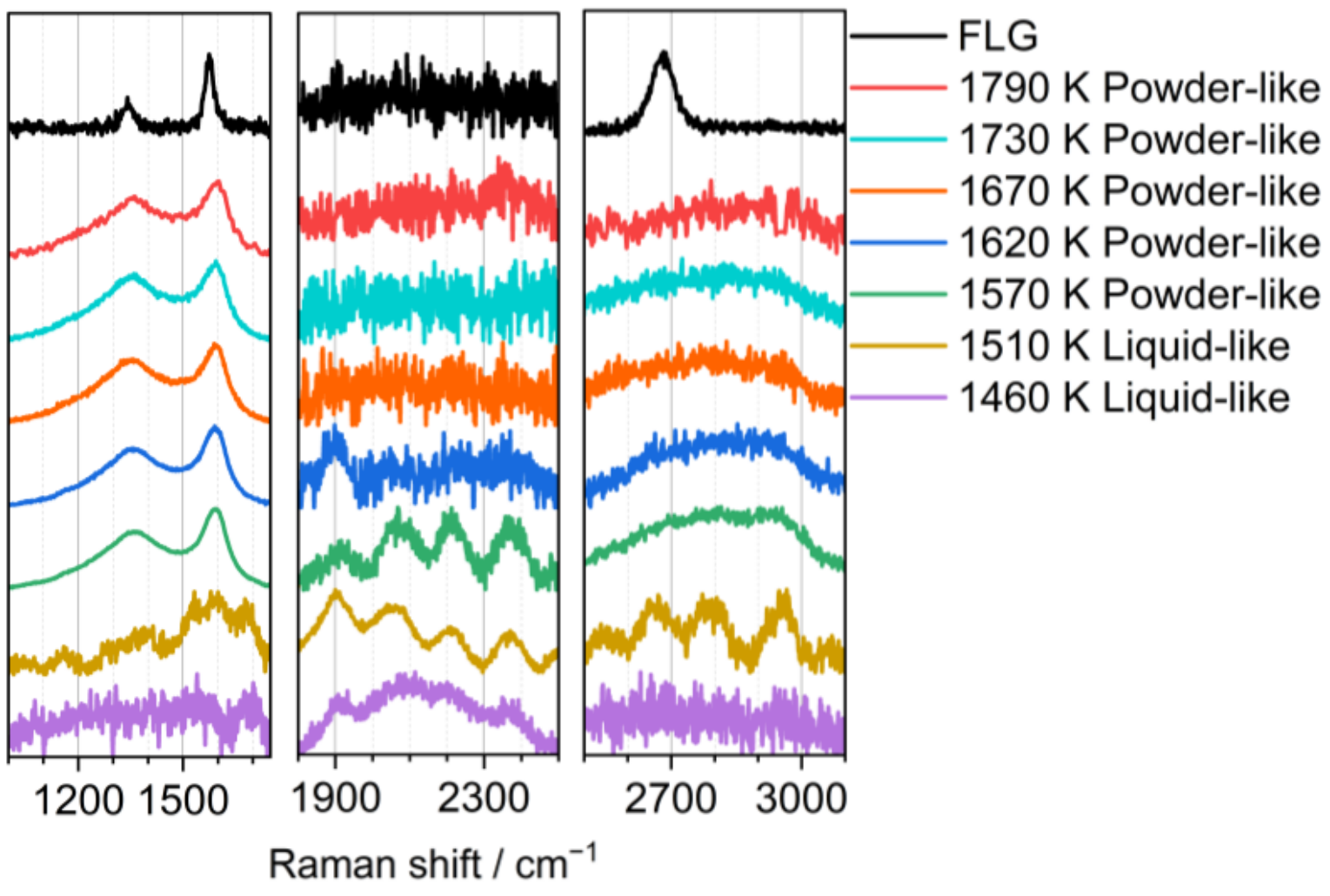

Fig. 6: Selected locally normalized Raman spectra of the collected samples for selected plateau temperatures (rounded to the nearest 10 K). D (~1350 cm⁻¹) and G (~1580 cm⁻¹) bands appear in the left panel at temperatures of 1570 K and above.

Figure 7 shows locally normalized FTIR spectra, while Table 2 represents the corresponding spectral features. A combination of the oop bands at 700–900 cm⁻¹ with an envelope at 1600 cm⁻¹ (C=C stretch) in addition to the non-separated collection of peaks at 3000–3100 cm⁻¹ (=C–H stretch) are three features assigned to aromatic groups that are used to identify aromatics over olefinic materials. On the other hand, a shoulder at 3000–3100 cm⁻¹ and the extension of the envelope at ~1600 cm⁻¹ to 1640–1670 cm⁻¹ partially opposes such an assignment, weakly signaling the partial presence of olefins at these low-temperature cases. However, these features, although already weak, deteriorate with temperature, indicating the disappearance of olefins before approaching the phase-limiting temperature. Accordingly, we assign the 3000–3100 $cm^{-1}$ region mainly to aromatic C–H in all experiments [50].

The phase transition discussed above was also seen in the FTIR signals. The peak at 1600 $cm^{-1}$ (aromatic C=C stretch) broadens in samples generated above 1570 K, a behavior widely reported for chars and soot undergoing aromatization [51, 52]. The maxima within the C=C envelope at 1550–1670 $cm^{-1}$ also resemble a temporal red-shift in the mid-temperature range. As shown later by deconvolution, the shifting trend originates from the vanishing of side-chain C=C (~1625 $cm^{-1}$) together with the growth of in-ring aromatic C=C (~1585 $cm^{-1}$).

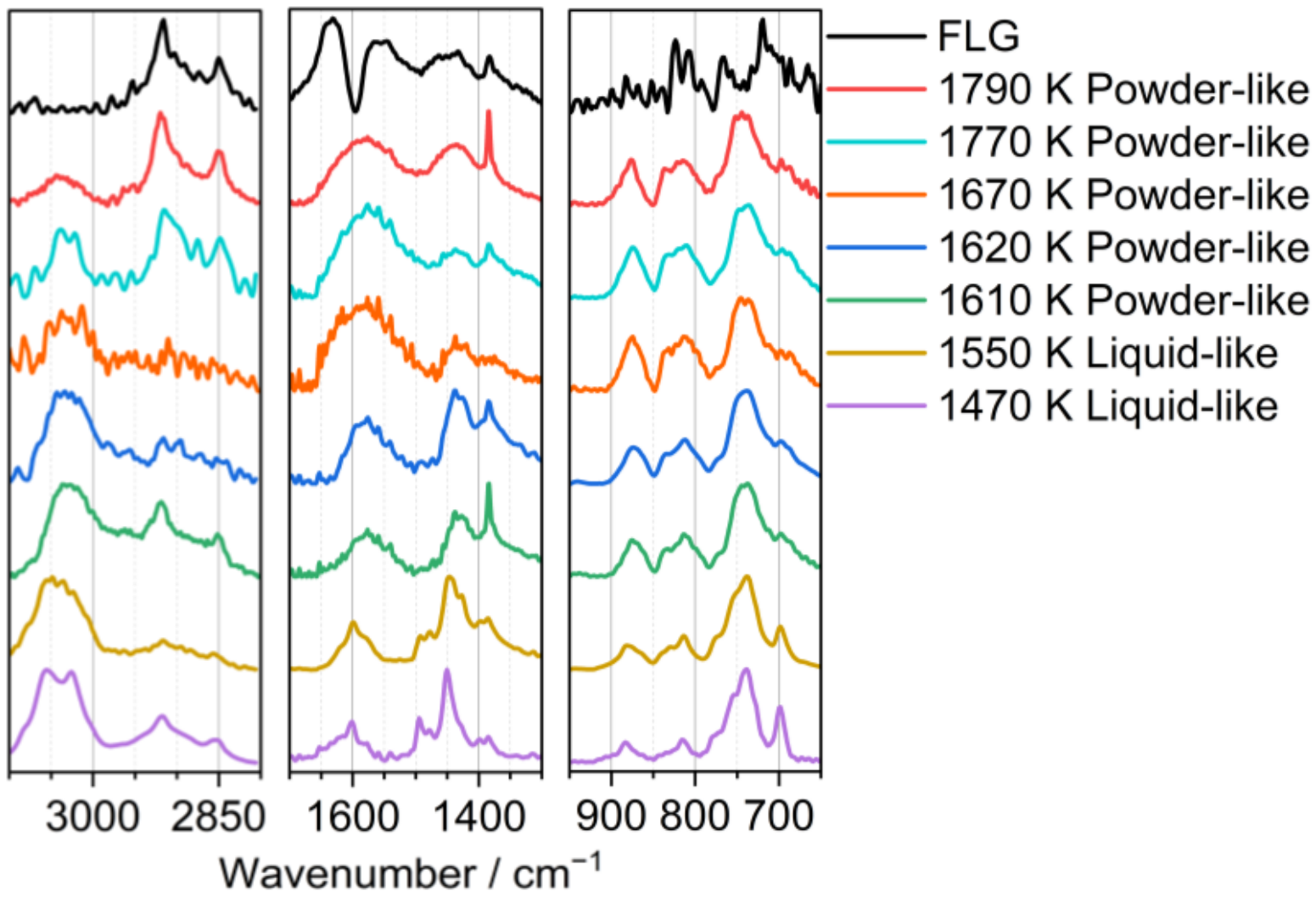

Fig. 7: Locally normalized FTIR spectra of samples at the range of plateau temperature under study (rounded to the nearest 10 K).

Additionally, a clear trend can be observed in the CH-stretch region at 1670 K and above, where the aromatic C–H stretch at 3000–3100 $cm^{-1}$ start to fade, whereas the peaks assigned to aliphatic C–H stretch at 2800–3000 $cm^{-1}$ are enhanced in agreement with previous FTIR studies on flame soot [53, 54]. In the FLG spectrum, no peaks appear in 3000–3100 $cm^{-1}$, which implies the low abundance of aromatic C–H bonds alongside a large peak at 1540–1600 $cm^{-1}$ (C=C) indicating the presence of a large $sp^2$ network. The broadening of the C=C envelope towards 1640 $cm^{-1}$ in the high-temperature cases or the distinguished peaks at this wavenumber for FLG is unlikely to mark the presence of C=C side chains since no signals within the 3000–3100 $cm^{-1}$ range are observable.

Deconvolution of the Raman and FTIR spectra (as a multi-band fitting procedure described in the supplementary material) was performed to scrutinize the features of the various molecular structures and to correlate them to chemical and structural interpretations for all temperatures investigated in this study. In Figures 8 and 9, the deconvolved peaks are shown for Raman and FTIR spectra, respectively. Deconvolution of the FTIR spectra in the C=C region enables us to represent normalized band areas using the in-ring aromatic C=C, rather than the area of the total C=C envelope that was used in previous studies [18, 55, 56].

Table 2: FTIR peak assignment.

| **Peak position / $cm^{-1}$** | **Assignment** |
|---|---|
| 700–690 | Quintet aromatic C–H oop bending (five adjacent C–H) [22, 23]. |
| 742 | Quartet aromatic C–H oop bending (four adjacent C–H) [22, 23]. |
| 778 | Trio aromatic C–H oop bending (three adjacent C–H) [22, 23]. |
| 835–815 | Duo aromatic C–H oop bending (two adjacent C–H) [22, 23]. |
| 886 | Solo aromatic C–H oop bending (isolated C–H) [22, 23]. |
| 1380 | Methyl ($–CH_3$) symmetric bending mode [17, 57]. |
| 1450 | Methyl ($–CH_3$). A component of the 1450 $cm^{-1}$ envelope: $CH_3$ asymmetric bending [17]. |
| 1460 | Methylene ($–CH_2–$). A component of the 1450 $cm^{-1}$ envelope: $CH_2$ scissoring [58]. |
| 1580 | Shoulder of the in-ring aromatic C=C envelope resolved in this study at 1580 $cm^{-1}$ [50, 57, 58]. |
| 1600 | Second peak of the in-ring aromatic C=C envelope resolved in this study at 1600 $cm^{-1}$ [50, 57, 58]. |
| 1620–1640 | Ring-conjugated C=C side-chains or alkene double bonds conjugated with aromatic rings (dienes, α,β-unsaturated carbonyls, styrenes), enhanced alkene band [50, 58]. |
| 1640–1680 | Isolated (un-conjugated) C=C side-chains, including terminal $CH_2$=CH– alkenes [50, 58]. |
| ~2890 | Aliphatic methine (tertiary CH) C–H stretch mode [17, 58]. |
| ~2915 | Aliphatic $CH_2$ asymmetric C–H stretch mode [17, 58]. |
| ~2945 | Aliphatic $CH_3$ asymmetric C–H stretch mode [17, 58]. |
| ~3022 | Aromatic =C–H stretch of $sp^2$ rings in the aromatic C–H region [17, 35, 36, 54]. |
| ~3055 | Aromatic =C–H stretch of $sp^2$ rings in the aromatic C–H region [17, 35, 36, 54]. |

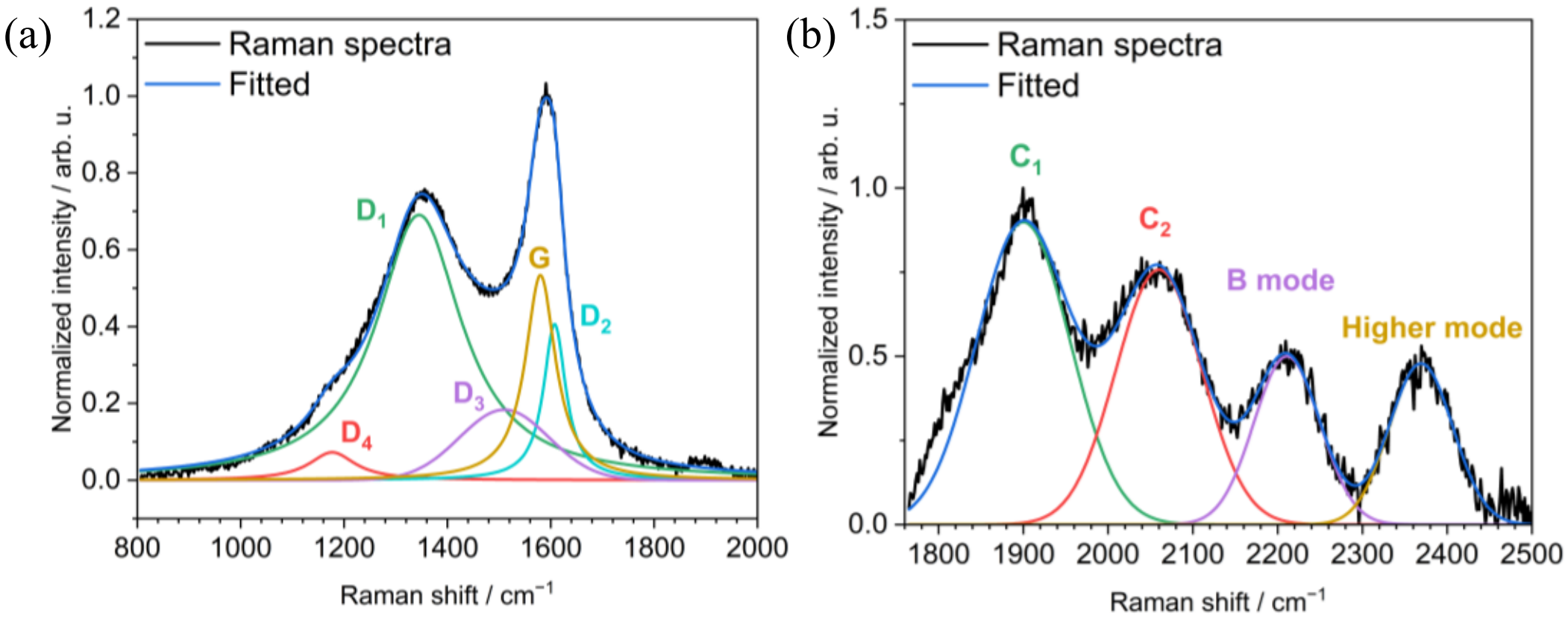

Fig. 8: Illustration of deconvolved Raman spectra and the resulting subpeaks for 1620 K (a), and 1510 K (b) cases.

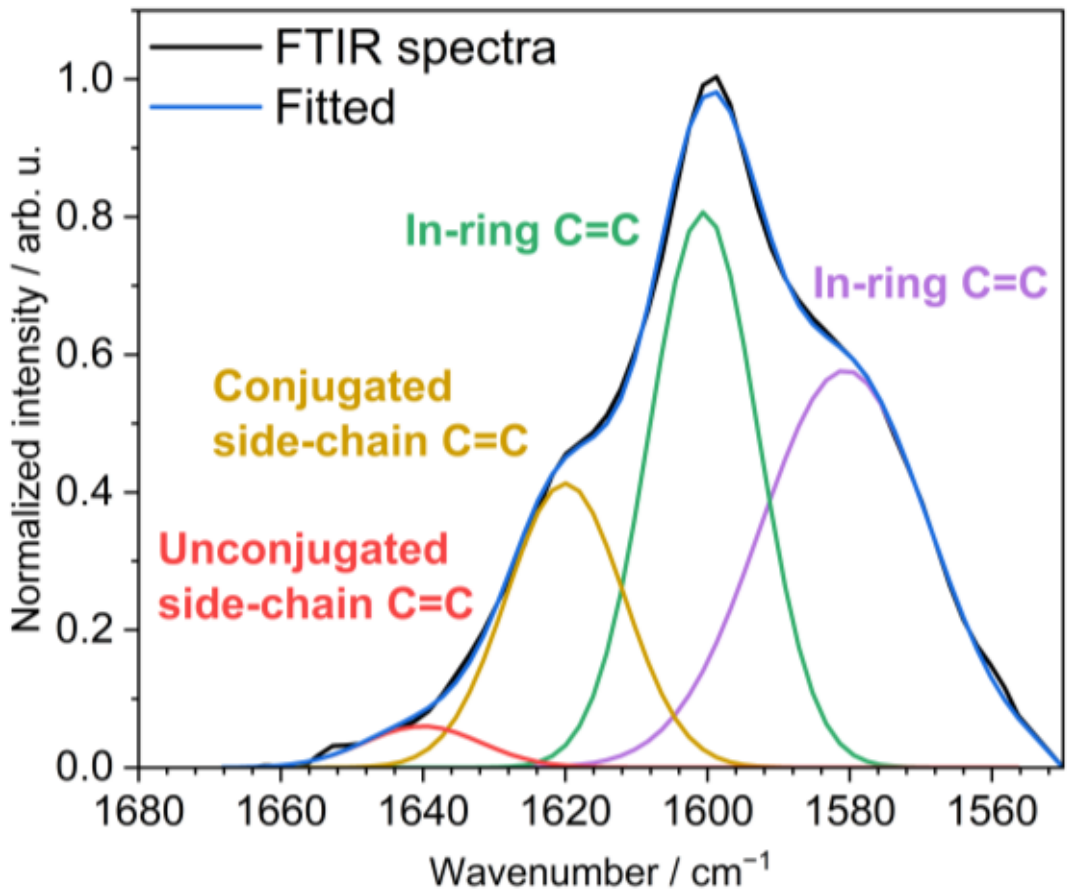

Fig. 9: Deconvolution of FTIR spectra in the C=C region for the 1620 K case

Figure 10 displays the temperature dependence of the first-order Raman ratios ($D_1/G$, $D_2/G$, $D_3/G$, and $D_4/G$) taken from the corresponding deconvolved sub-peak areas. The values are averaged over three measurements on samples collected from different spots on the same filter (same experiment). The $D_1/G$ has a local maximum near 1630 K, reflecting increased edge-defect formation after the phase-limiting stage, followed by a decrease in the ratio to 1670 K, and a sharp rise to its overall maximum around 1730 K. According to the Tuinstra–Koenig [59] and Ferrari–Robertson [25] framework, this ratio is inversely proportional to the in-plane crystallite size, known as lateral area (La). A minimum in $D_1/G$ means that La of the $sp^2$ surface has reached a maximum. The valley in the $D_1/G$ ratio around 1670 K therefore represents a peak in the in-plane crystallite size, which coincides with the peak in $d_p$ showed in Fig. 5. The $D_2/G$ ratio exhibits a similar trend to the $D_1/G$ ratio in terms of a minimum at 1670 and a maximum at 1730 K. An increase in $D_2/G$ indicates an enhanced structural disorder of perturbed turbostratic/stacked layers. Thus, a minimum in misoriented stacked layers is observed around 1670 K. The $D_3/G$ ratio, which tracks amorphous $sp^2$ carbon, decreases from 1620 K to 1670 K, which indicates an increase in ordering level from amorphous structures toward more ordered $sp^2$ domains (more organized graphene-like aromatic domains). This ratio approaches zero at higher temperatures and maturation, indicating the absence of amorphous $sp^2$ carbon. The $D_4/G$ ratio, associated with $sp^3$/polyenic fragments at ring edges, shows little variation at temperatures below 1670 K, a sharp increase to a maximum at 1730 K, followed by a decrease at higher temperatures, confirming the loss of these features at higher levels of maturity.

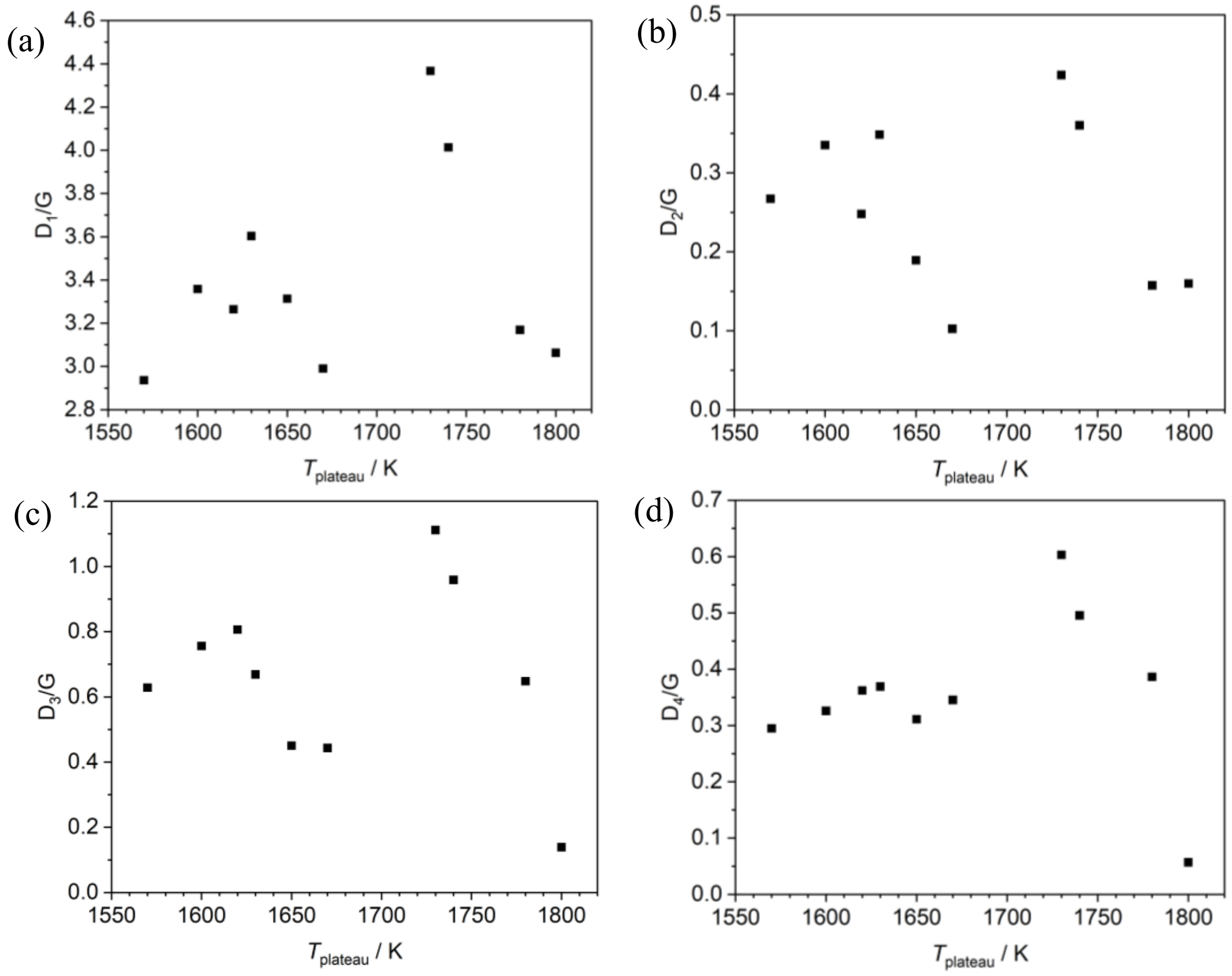

Fig. 10: Raman peak area ratios of (a) $D_1$/G, (b) $D_2$/G, (c) $D_3$/G, and (d) $D_4$/G at different plateau temperatures

Combining these results, observations of a peak in the $d_p$ curve and transition of the functional groups tracked by FTIR spectra with temperature, demonstrate a pivotal change at around ~1670 K. The Raman results demonstrated sudden changes as a drop in edge defects and amorphous carbon. In combination with a drop in the misoriented stacked graphitic layers at around 1670 K, this temperature can be assigned as the "ordering threshold" in this study. At higher temperatures, a secondary maximum in $D_1$/G, $D_2$/G, $D_3$/G and $D_4$/G near ~1730 K might indicate renewed defect and stacking-fault formation due to fast growth/surface addition. The fast growth in this temperature range is confirmed by the sharp rise of the optical density in this temperature range (Fig. 2), consistent with rapid particle growth, introducing disorder faster than thermal restructuring can reduce disordered structures within the ~2 ms plateau temperature time.

Figure 11 shows the Raman ratios used to track the evolution of sp-hybridized chains. The $C_1$/G and $C_2$/G ratios, representing $sp^1$-chains, decrease sharply between 1500 K and the phase-limiting temperature (1570 K), showing that both long polyynes ($C_1$) and shorter cumulenes ($C_2$) are rapidly decreasing as the first $sp^2$ network condenses. As the direct $C_1/C_2$ ratio is independent of the Raman G band, it can be tracked to lower temperatures. The $C_1/C_2$ ratio undergoes a significant increase in which long polyynes are rapidly fragmented/converted into shorter cumulenes, reaching a maximum just beyond the phase-limiting temperature. Close to the ordering threshold, $sp^1$-chain signals approach zero, indicating that edge-linked $sp^1$ bridges (polyynes and polycomulenes) are disappearing and do not contribute to the subsequent growth.

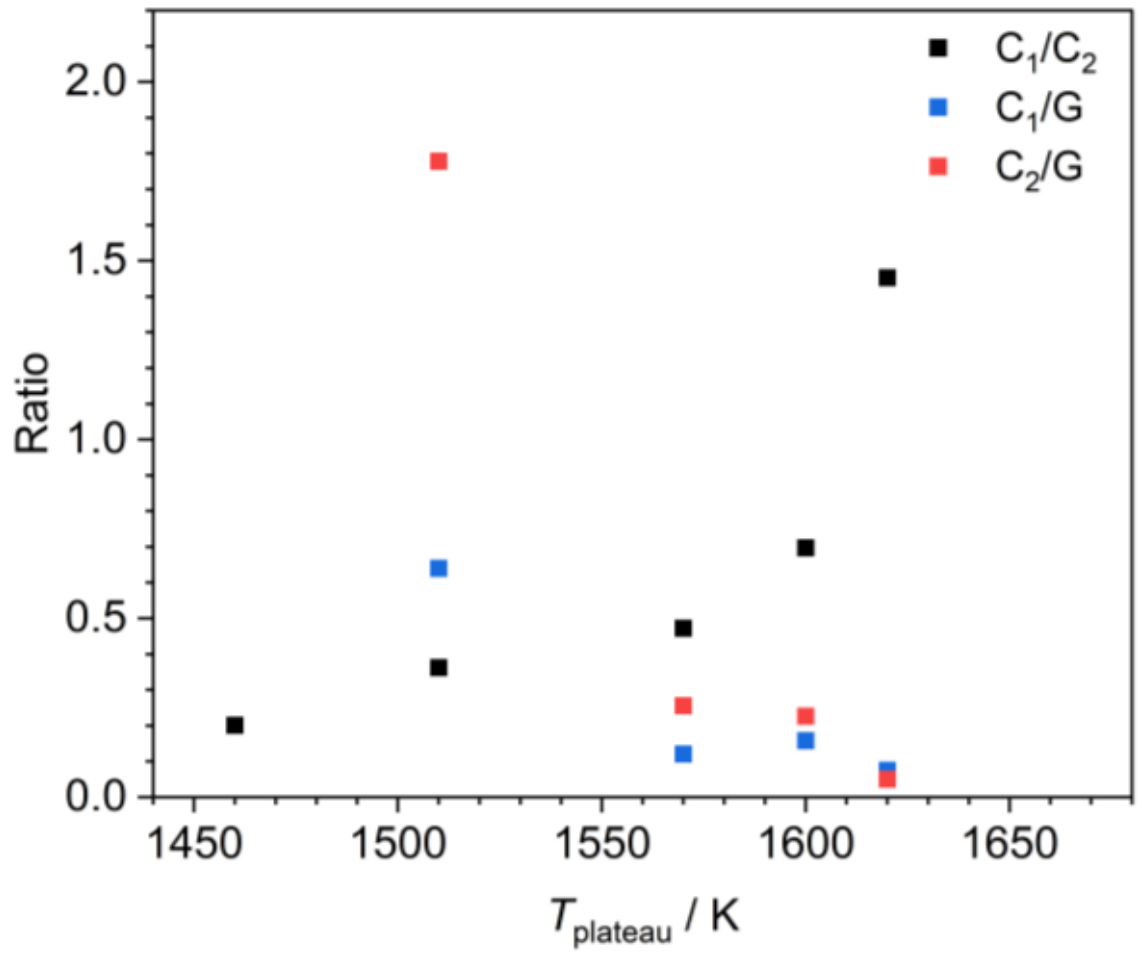

Fig. 11: Raman peak ratios for sp-chain ($C_1$: polycumulenes, and $C_2$: polyynes) bands.

Figure 12 demonstrates the ratios of the side-chain C=C normalized by the in-ring C=C FTIR peak area versus temperature. Both ratios are large at 1450–1480 K, consistent with a side-chain-rich, liquid-like nascent soot. Close to the phase-limiting temperature (~1570 K), both unconjugated and conjugated C=C ratios approache zero. This suggests the completion of a gradual migration of π-conjugation (i.e., delocalization of π electrons over adjacent $sp^2$ sites/fused rings, producing resonance stabilization) from conjugated C=C side-chains into the $sp^2$ core and the end of π-network-driven edge growth by the phase-limiting temperature through those species.

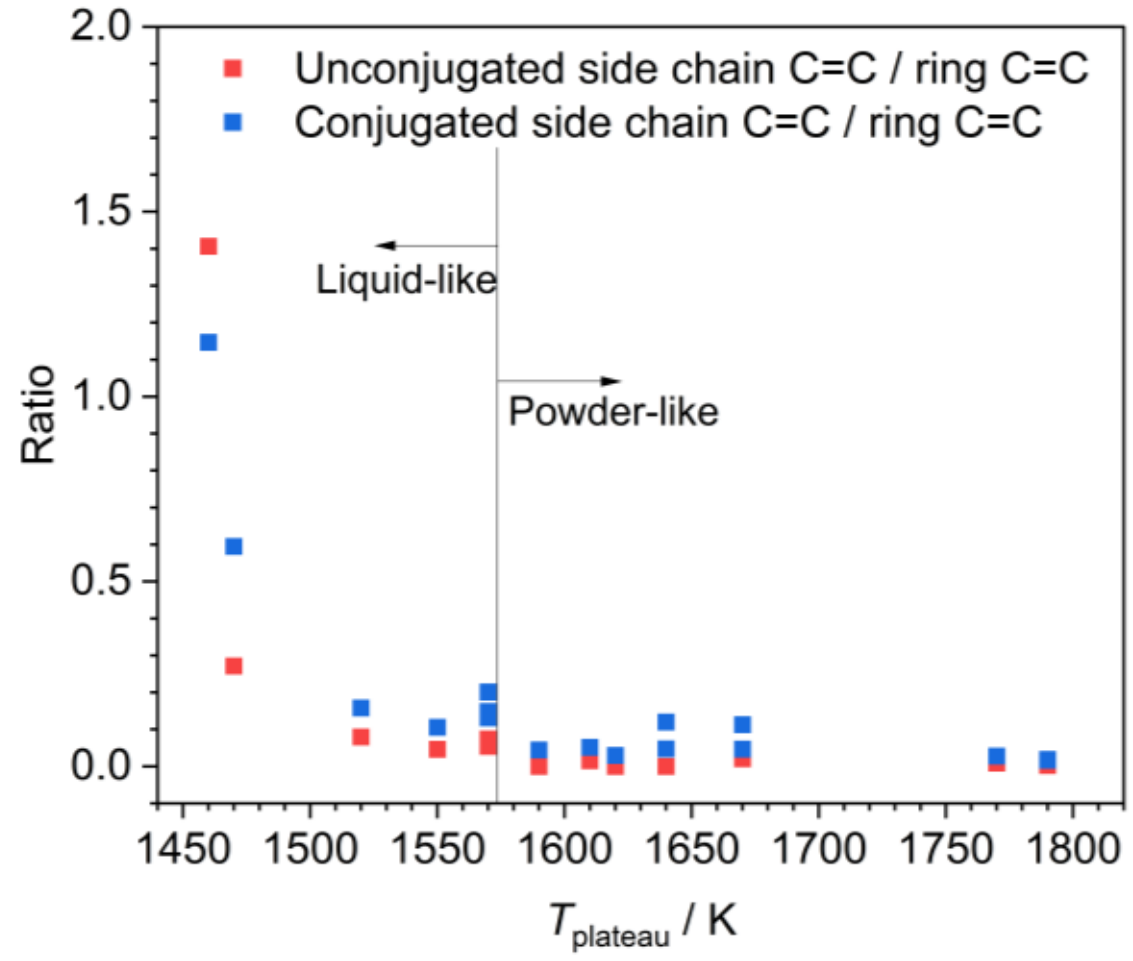

Fig. 12. Temperature-dependent side-chain C=C stretch FTIR signal normalized by in-ring C=C signal.

Figure 13 presents the ratios of C–H stretching relative to the in-ring C=C stretch for aliphatics and aromatics, determined from FTIR spectra. Aromatics are predominant at the lower temperatures, highlighting the PAH-rich nature of the nascent, liquid-like material. We attribute the changes in this ratio below the phase-limiting temperature to the vanishing olefins as well as to the inseparability of olefins and aromatic C–H stretch peaks at 3000–3100 $cm^{-1}$. The ratios of both aromatics and aliphatics experience a change in their rate of decrease at the phase-limiting temperature, with little change at higher temperatures. Notably, the trend for the ratio of aliphatics to C=C shown here differs from that reported by Cain et al. [18] for micro-FTIR analyses of flame-generated soot sampled at different heights above the burner. As they showed an increasing trend for the aliphatics over the C=C ratio versus temperature, this might mark a difference between soot generated flames (in the presence of oxidizing species) and pyrolysis in shock tubes.

Moving to lower wavenumbers, the contribution of topology/structure is enhanced in the emerging peaks. Associating the FTIR peak area ratios with edge structures enables us to recognize additional characteristics of these evolving edge structures. Particular edge structures are potential sites for hosting localized or delocalized unpaired electrons. As de-localized electrons spread over the aromatic $sp^2$ network and play a role in preventing them from re-fragmentation by heat, the localized electrons influence the reactivity.

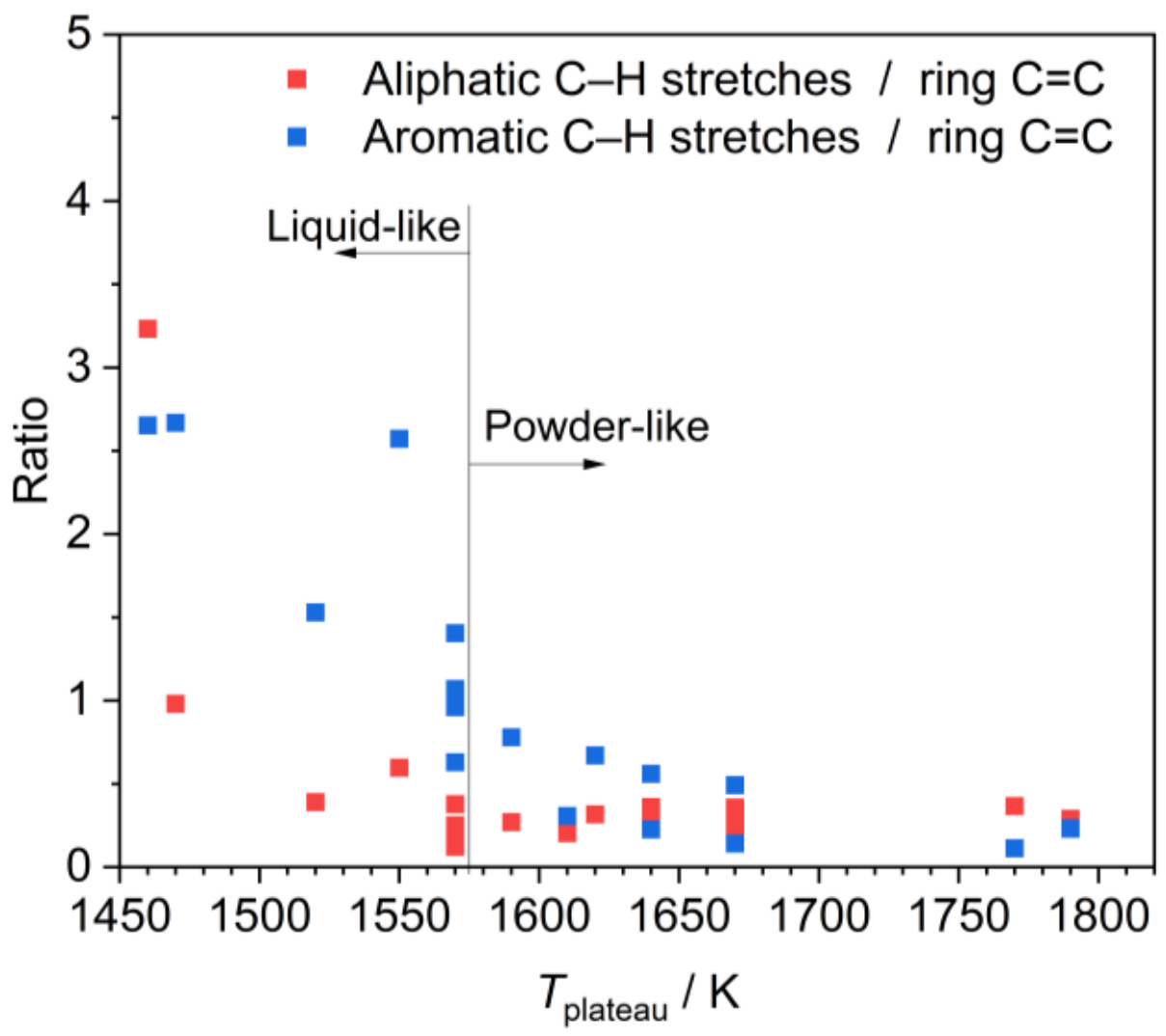

Fig. 13: Temperature-dependent aromatic and aliphatic C–H stretch FTIR signal normalized by in-ring C=C signal.

Figure 14 shows a schematic of the introduced edge structures related to aromatic rings according to the assignments in Table 2. The presence of quartets and trios around six-membered rings can create bay pockets [60] and is indirectly entangled with edge defects that lead to the formation of five-membered rings [61, 62]. Both edge structures represent potential sites for localized unpaired electrons [12, 14]. Additionally, isolated hydrogen atoms (solos) can create zig-zagged segments as potential sites for localized π-radicals [63-65]. On the other hand, duos can form armchair [20, 65] and K-regions [66] as potential sites for delocalized unpaired electrons [65]. In other words, whereas quartet-/trio- and solo-H structures by themselves promote localized unpaired electrons, the presence of duo -H sites provides the structural setting that favors delocalization sites. Moving away from the aromatic rings, aliphatic side-chains (determined by aliphatic bending modes detected in FTIR) have a high potential for hosting benzylic radicals. A benzylic radical is resonance-stabilized but only within the first one-and-a-half rings; thus, they can be termed partially localized and moderately reactive [67].

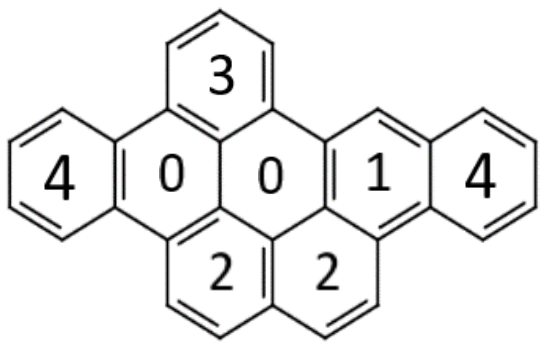

Fig. 14: Schematic structure with C–H out-of-plane vibrations represented by the number of adjacent H atoms around aromatic rings as solos (1), duos (2), trios (3), and quartets (4).

Figure 15 shows the ratio of the peak areas in the FTIR spectrum associated with the potential structures that can carry unpaired electrons. A higher number of quartets or trios at lower reaction temperatures can host an abundance of contiguous edge H, leading to the formation of bay pocket/five-membered rings, as potential sites for localized electrons. Additionally, a higher abundance of aliphatic bends represents the existence of benzylic sites that favor partially localized electrons. A stable trend observed in solos up to the phase-limiting temperature of 1570 K is followed by a steady decrease from 1630 K until 1670 K. This indicates a higher potential for the presence of zig-zagged edges as additional sites for electron localization before the phase-limiting temperature. Also, duos reach a maximum at the phase-limiting temperature, and decrease concurrently with solos from 1630 K up to 1670 K. The increase in duos indicates the emergence of armchair and K-regions, and subsequently the presence of thermally stable edge structures at the phase-limiting temperature. Across the phase-limiting temperature, as structures that could localize electrons decline, the structures that favor electron delocalization emerge and persist up to the ordering threshold. This observation may emphasize the influential role of radicals in soot inception [3, 12]. Interestingly, the armchair and K-region structures (inferred from the increase in duos from FTIR spectra) and the Raman edge-defect indicator $D_1/G$ both decrease from 1630 K until the ordering threshold (1670 K) is approached. This suggests a relation between the local maximum in the edge defect interpreted from Raman spectroscopy and hydrogen-terminated edge structures in this range of reaction temperature. In parallel, FTIR edge C–H features indicate that ring edges lose most of their hydrogen termination above the ordering threshold (~1670 K). Therefore, the pronounced second increase in $D_1/G$ at 1730 K, above the ordering threshold, cannot be attributed primarily to H-terminated ring-edge structures. This implies an increasing contribution from non-C–H-related sources of disorder (e.g., internal lattice defects).

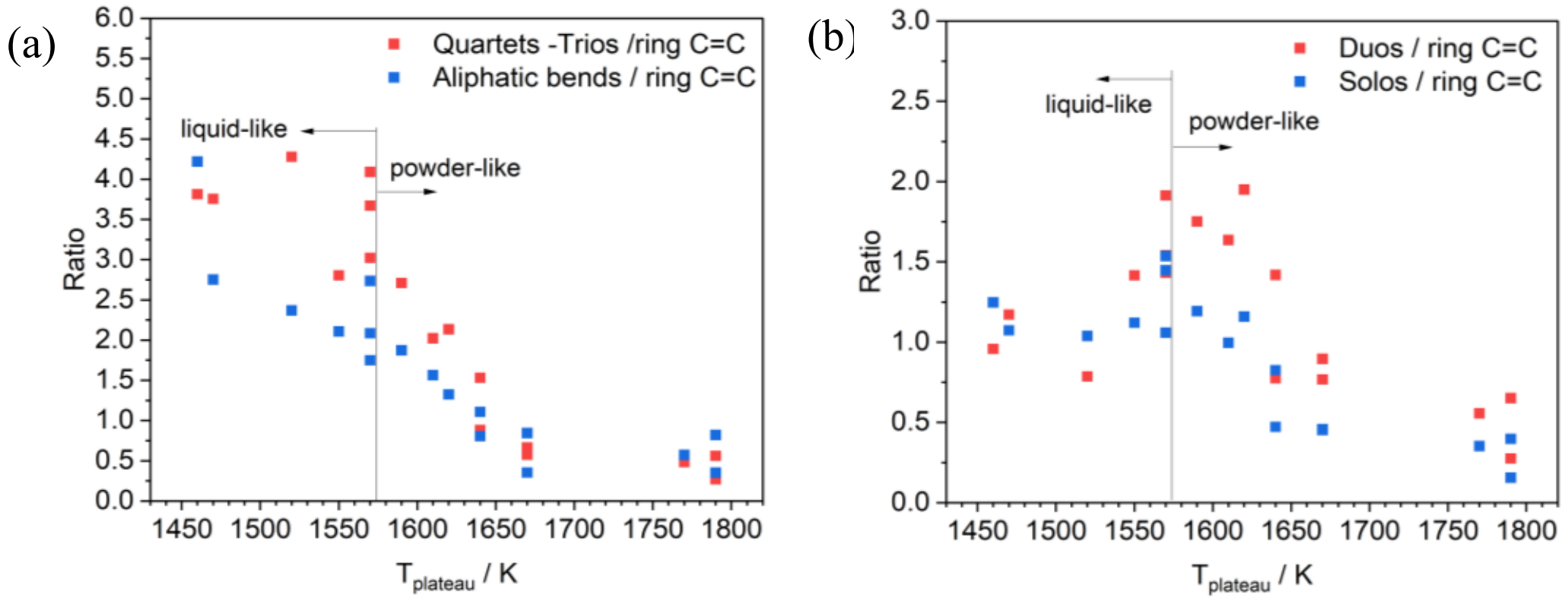

Fig. 15: Temperature-dependent C–H bending FTIR signals of solo, and duo structure (a), and Quartet, and aliphatic structures (b) normalized by in-ring C=C signal representing potential radical-hosting for selected plateau temperatures.

## 3. Conclusions

The primary goal of this study is to investigate the early stages of soot formation, especially the zone between the formation of liquid-like and solid carbonaceous nanomaterials, during pyrolysis of shock-heated toluene (2 % in argon). Experiments were conducted across an initial temperature ($T_5$) range of 1496–2079 K (related to plateau temperatures after initial pyrolysis of 1450–1800 K) at pressures of 2.0 ± 0.1 bar with a plateau temperature time of ~2 ms. *In situ* laser extinction measurements were conducted for each experiment, and particles were extracted from the product gas after cooling for TEM imaging, and Raman and FTIR spectroscopy, to study the primary particle size, structural order, and functional groups. Spectral deconvolution for Raman and FTIR was employed to identify individual contributions of the different vibrational modes. The main findings identified three distinct regions associated with the plateau temperature, a liquid-like region, a powder-like region with a high degree of disorder, and a powder-like region with more ordered structures. The boundaries for these regions and key characteristics are the following:

**Liquid-like samples towards the phase-limiting temperature (1450–1570 K):**

- Negligible *in situ* laser extinction is observed, while TEM images show some poorly-defined structures, and Raman spectra lack D or G bands, indicating no extended $sp^2$ order.
- A sharp decrease in polyynes and polycumulenes is observed from lower temperatures up to the temperature of 1570 K, consistent with breaking of sp chains that connect PAH structures.
- FTIR spectra are dominated by signals of aromatics and methyl C–H stretch modes, the collapse of the side-chain C=C, and a higher probability of edge structures such as bay pockets, and benzylic positions, pointing to a PAH-rich, potentially radical-dense liquid core.

- An increase in optical density, the first appearance of Raman D and G bands, and particle morphologies observed by TEM, together mark 1570 K as the phase-limiting temperature.
- At this temperature, the emergence of potential structures for electron delocalization, and thermal stability emphasizes the potential role of the radical in the inception.

**Powder-like samples in the phase-limiting temperature range towards the ordering threshold (1570–1670 K):**

- Decay of ring-conjugated C=C functional groups to ~0 is observed, as an indication of the completion of a gradual migration of π-conjugation from C=C side-chains into the $sp^2$ core. This migration can mark this stage as the end of π-network-driven edge growth via side chain C=C.
- Optical density sharply increases near 1640 K, indicating the initiation of fast growth of solid soot particles.
- At the ordering threshold (~1670 K), while the primary particle diameters demonstrated a maximum, the $D_1/G$ and $D_2/G$ ratios related to the edge defects, graphitic surface disorder, and misoriented stacked graphitic layers decrease, indicating the emergence of an ordered structure. Additionally, the amorphous $sp^2$ ($D_3/G$) decreased, further indication of a higher level of ordering.

**Powder-like samples beyond the ordering threshold (1670–1750 K):**

- *In-situ* extinction measurements demonstrate a continued increase in optical density with temperature in this interval. All FTIR-related signal ratios, become temperature independent. The ratios representing defects and disorder reach a maximum around 1730 K which is followed by a flat region for $D_1/G$ and $D_2/G$, while $D_3/G$ and $D_4/G$ approach zero, consistent with rapid growth introducing disorder faster than thermal restructuring can reduce disordered structures within the ~2 ms plateau time.
- FTIR edge-hydrogen features collapse while the Raman-detected edge-defect ($D_1/G$) exhibits a second maximum. This decoupling suggests that the high-temperature (1730 K) $D_1/G$ increase is not primarily governed by H-terminated ring-edge structures.

Building on this framework, future work will apply electron-paramagnetic resonance (EPR) spectroscopy with the same shock-tube-sampling protocol, enabling direct study of radical characteristics and their transitions during particle maturation, further refining our understanding of radical-mediated soot formation and growth.

**CrediT authorship contribution statement**

**Meysam K. Rezaeian:** Writing – original draft, Investigation, Data curation, Experiments.
**Can Shao:** Methodology, Investigation, Data curation, Writing-review & editing.
**Jürgen Herzler:** Writing-review & editing, Data curation.
**Mustapha Fikri:** Writing-review & editing, Data curation, Supervision, Resources, Project administration.
**Greg J. Smallwood:** Supervision, Writing-review & editing, Data curation, Conceptualization.
**Christof Schulz:** Writing – review & editing, Data curation, Supervision, Resources, Project administration, Funding acquisition, Conceptualization.

**Declaration of competing interest**

The authors declare that they have no known competing financial interests or personal relationships that could have appeared to influence the work reported in this paper.

**Acknowledgement**
We acknowledge financial support by the German Research Foundation (DFG 491110473) that also supports Greg Smallwood as Mercator Fellow. Can Shao acknowledges financial support by the Alexander von Humboldt Foundation. The authors appreciate the invaluable technical support of Ludger Jerig and Birgit Nelius during the experimental measurements. Materials characterization by TEM and Raman were carried out at the Interdisciplinary Center for Analytics on the Nanoscale, ICAN. Their careful preparation, operation, and maintenance of the equipment were essential for the successful completion of this work.

# Supplementary Material

# Effect of reaction temperature on nascent carbonaceous particles from toluene shock-tube pyrolysis: Insights from FTIR and Raman spectroscopy

Meysam K. Rezaeian[a], Can Shao[a*], Jürgen Herzler[a], Mustapha Fikri[a], Greg J. Smallwood[b], Christof Schulz[a]

[a]*EMPI, Institute for Energy and Materials Processes – Reactive Fluids and CENIDE, Center for Nanointegration Duisburg-Essen, University of Duisburg-Essen,47048 Duisburg, Germany*
[b]*Metrology Research Centre, National Research Council Canada, Ottawa, ON K1A 0R6, Canada*

*Corresponding author.

## Temperature variation as a result of toluene pyrolysis

Figure S1 shows the deviation of plateau temperature from $T_5$ as a function of the thermochemistry of toluene pyrolysis. The 3 % and 15-K error bars are applied to the plateau temperature and $T_5$, respectively.

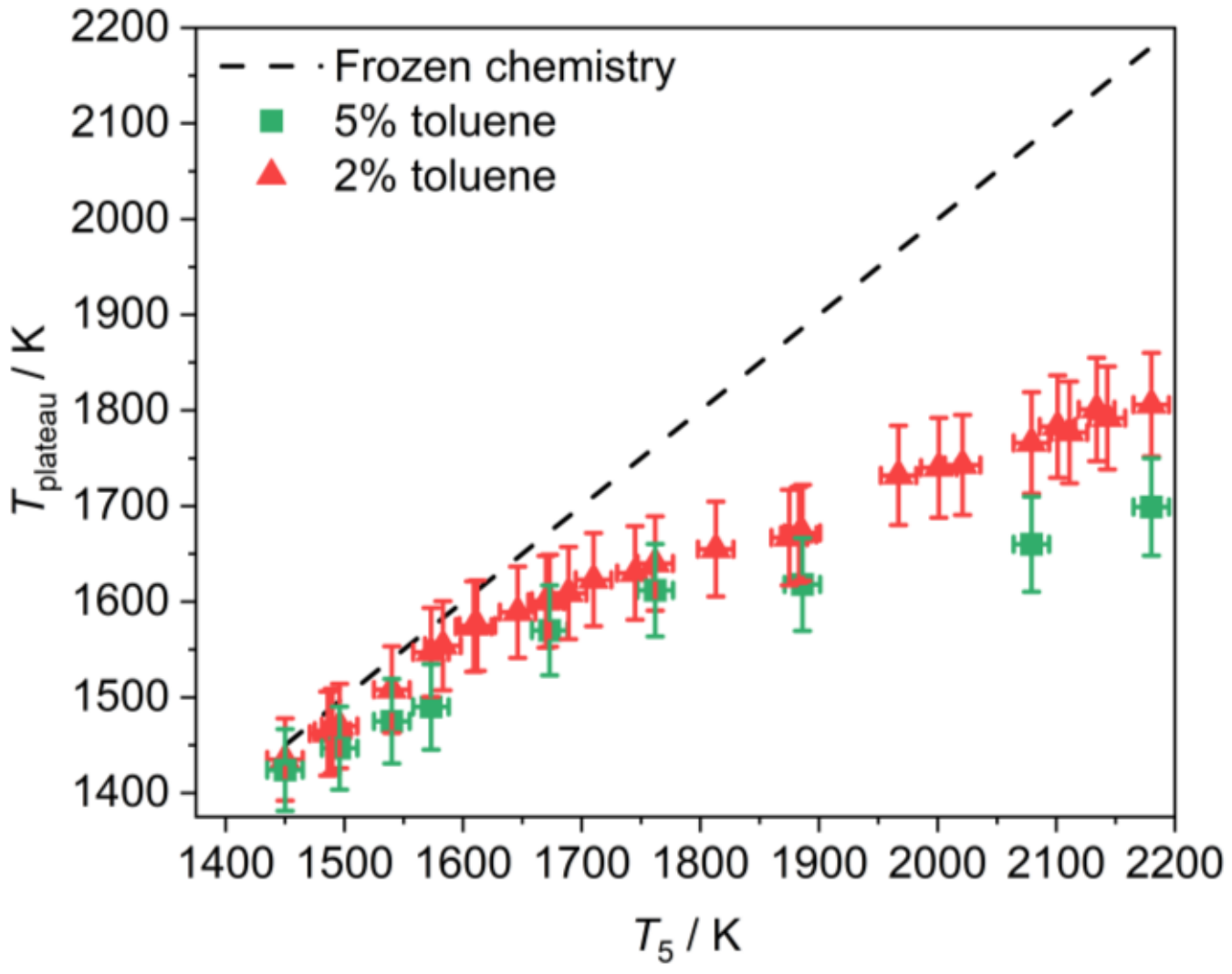

Fig. S1: Plateau temperature versus $T_5$ for different toluene concentration.

## Raman measurement details

Figure S2 compares Raman spectra for a collection of materials, generated from the shock tube under varying conditions in addition to the clean filter. The signal measured on the clean filter confirms that there are no interfering features in the spectrum obtained at 1510 K, where the liquid-like particles were left on the filter for Raman analysis.

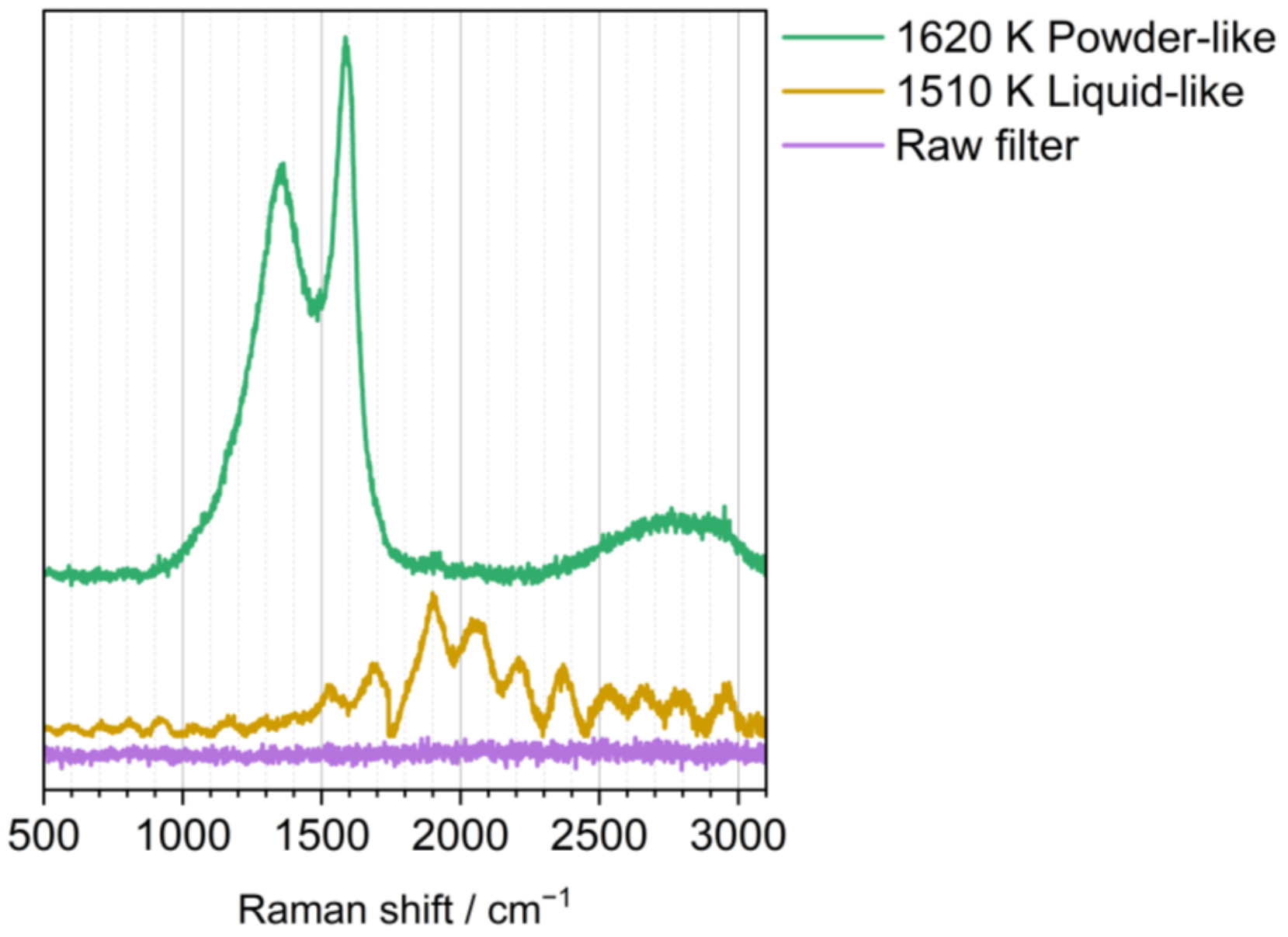

Fig. S2: Raman spectra for selected shock-tube-generated materials samples at the given different plateau temperatures (rounded to the nearest 10 K), compared to the signal generated by measurements on the clean filter.

## Raman, and FTIR spectra deconvolution

Signal ratios are derived from areas under each peak. For convolved regions, the sub-peak areas are calculated after nonlinear least-squares fitting [1]. The least-squares fitting describes each spectrum as a sum of individual sub-peaks parameterized by peak height, center position, and full width at half maximum (FWHM). The fitting process incorporates an optimization, with constraints enforcing positive peak heights and center positions within a prescribed tolerance. The non-linear least-squares fit is considered converged when further iterations changed *i*) the cost function (sum of squared residuals), *ii*) the fitted peak parameters (heights, positions, widths), or *iii*) the gradient norm, by less than $1\times10^{-8}$. The criterion for the fitted peak parameters (heights, positions, widths) is denoted by the tolerance, XTOL. The goodness-of-fit is evaluated with the coefficient of determination ($R^2$) computed on the baseline-corrected spectrum. Considering the convergence criterion mentioned above, for most cases $R^2 \geqslant 0.994$ is achieved. Additionally, the ratio of the root mean square of residual over the spectrum maximum, is used as a parameter ($RMS_{rel}$), which is below 1.9 % for most cases.

A sensitivity analysis for spectral features with respect to the variation of the XTOL is carried out. Sensitivity indicators are defined as below:

$$(\delta_{\rm abs})_T^{\rm XTOL} = (Q)_T^{\rm XTOL} - (Q)_T^{\rm ref} \tag{1}$$

$$(\delta_{\rm rel})_T^{\rm XTOL} = \frac{\left|(\delta_{\rm abs})_T^{\rm XTOL}\right|}{(Q)_T^{\rm ref}} \quad , \tag{2}$$

where $Q$ is a quality factor that holds for $D_1/G$, $D_2/G$, $D_3/G$, and $D_4/G$ ratios, in addition to $R^2$, and residual $RMS_{rel}$. The indices ref refer to case with XTOL equal to $1\times10^{-8}$. It is demonstrated below that $\delta_{\rm rel}$ is altered by less than 12 % at the fit quality used here.

The deconvolution results are sensitive to lineshapes but settled for $D_1$, $D_2$, $D_3$, $D_4$, and G in the literature [2, 3] (as all Lorentzian except Gaussian for $D_3$), which are thus treated as fixed conditions. For $C_1$ and $C_2$, although Gaussian line shapes are used as suggested by Le et al. [4], the $C_1/G$ and $C_2/G$ trends include sharp gradients that are unlikely to change if different lineshapes are used. All subpeaks for deconvolution of the C=C region in the FTIR spectra are assumed to be Gaussian. The deconvolved in-ring C=C peaks are summed up (two out of four in Fig. 9) and subsequently make the most contribution to the region. The overlap of the side chain C=C sub-peaks with the in-ring C=C is not significant, thus switching to Lorentzian lineshapes resulted in similar qualitative interpretations.

Figure S3 shows the ratios of the subpeaks for $D_1$, $D_2$, $D_3$, and $D_4$ over G against the plateau temperature. Two criteria are presented as the goodness-of-fit ($R^2$) and relative root mean square ($RMS_{rel}$) of the deconvolution residuals to present the fit quality.

$$RMS_{rel} = \text{residual RMS / Maximum of baseline-corrected spectrum} \tag{3}$$

By modifying XTOL, different levels of accuracy are achieved. Each individual curve in Fig. S4 is plotted with different levels of fitting accuracy in accordance with different XTOLs presented in Table S1. For cases with XTOL= $10^{-5}$ and $10^{-8}$, the curves are identical. This demonstrates that the fitting converged to a unique solution. Figure S4 represents the converged criteria based on the data in Table S1. To calculate the points, the values are averaged over the range of reaction plateau temperatures above the phase transition (1570 to 1800 K).

Table S1: $R^2$, and residual $RMS_{rel}$ averaged over the range of reaction plateau temperatures for different deconvolution tolerances (XTOL).

| XTOL | $R^2$ mean over reaction plateau temperature range | Residual $RMS_{rel}$ mean over reaction plateau temperature range |
|---|---|---|
| $2\times10^{-1}$ | 0.9819 | 0.0352 |
| $1\times10^{-1}$ | 0.9894 | 0.0273 |
| $5\times10^{-2}$ | 0.9924 | 0.0211 |
| $2\times10^{-2}$ | 0.9940 | 0.0191 |
| $1\times10^{-5}$ | 0.9943 | 0.0189 |
| $1\times10^{-8}$ | 0.9943 | 0.0189 |

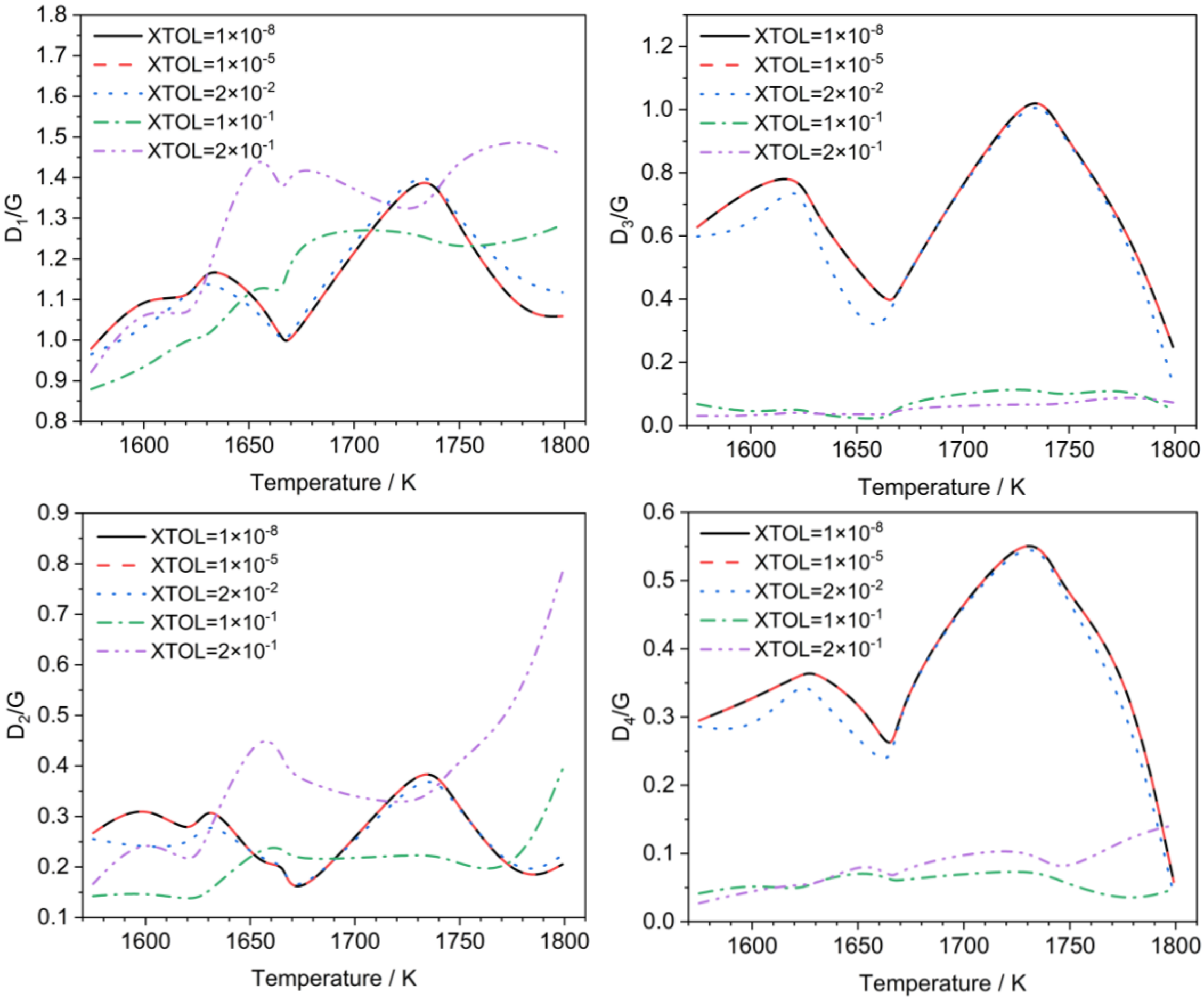

Fig. S3: Raman peak ratios $D_1$, $D_2$, $D_3$, and $D_4$ over G versus reaction plateau temperature for different levels of deconvolution fitting accuracy (XTOL).

For XTOL = 2×10$^{-2}$, the mean relative deviation of the band ratios remained below ~12 %, whereas the fit quality metrics changed negligibly. In contrast, for XTOL ≥ 1×10$^{-1}$, the band ratios became strongly sensitive to the tolerance (mean deviations ~50–100 %), even though $R^2$ still changed by less than ~1 %. With the nonlinear solver constrained to converged solutions (XTOL ≤ 1×10$^{-5}$, mean $R^2 > 0.994$ and mean residual $RMS_{rel} < 1.9$ %), the remaining sensitivity of the $Q$ ratios to XTOL is ~10 % on average (see Table S2). Under the converged and non-sensitive condition, the $D_3$ and $D_4$ contributions approach zero at the highest temperatures, consistent with the observed spectra and supporting the physical correctness of the fitted trends. Figure S5 shows the mean $\delta_{\mathrm{rel}}$ for each parameter under sensitivity analysis.

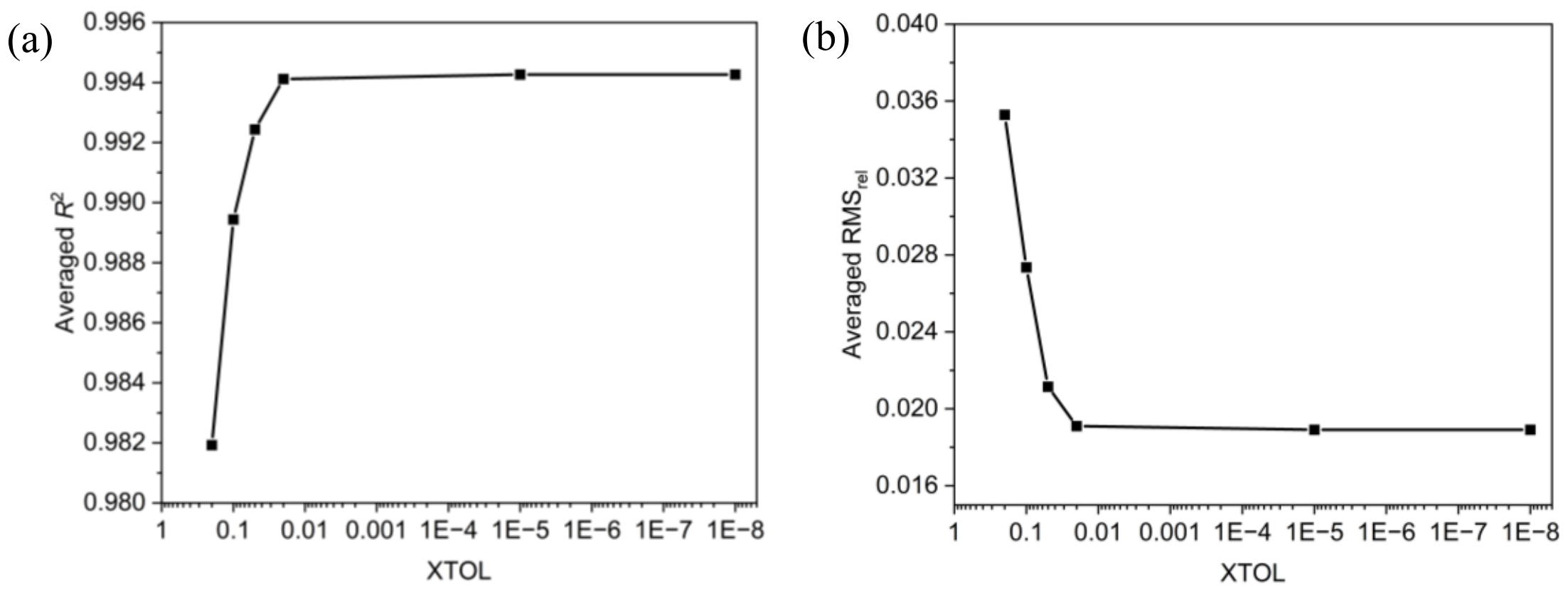

Fig. S4: (a) Averaged $R^2$, and (b) residual $RMS_{rel}$ over temperature with respect to XTOL.

Table S2: Sensitivity parameters of Raman subpeak ratios, $R^2$, and residual $RMS_{rel}$ at different levels of deconvolution fitting accuracy.

| | XTOL | $\delta_{rel}$ average | max $\delta_{rel}$ |
|---|---|---|---|
| $D_1/G$ | $1\times10^{-5}$ | 0 | 0 |
| | $2\times10^{-2}$ | 0.0341 | 0.0849 |
| | $1\times10^{-1}$ | 0.1408 | 0.2821 |
| | $2\times10^{-1}$ | 0.2153 | 0.4849 |
| $D_2/G$ | $1\times10^{-5}$ | 0 | 0 |
| | $2\times10^{-2}$ | 0.0885 | 0.2931 |
| | $1\times10^{-1}$ | 0.4810 | 1.0370 |
| | $2\times10^{-1}$ | 1.0343 | 2.8186 |
| $D_3/G$ | $1\times10^{-5}$ | 0 | 0 |
| | $2\times10^{-2}$ | 0.1216 | 0.4757 |
| | $1\times10^{-1}$ | 0.8939 | 0.9599 |
| | $2\times10^{-1}$ | 0.9030 | 0.9607 |
| $D_4/G$ | $1\times10^{-5}$ | 0 | 0 |
| | $2\times10^{-2}$ | 0.0918 | 0.3241 |
| | $1\times10^{-1}$ | 0.7802 | 0.9309 |
| | $2\times10^{-1}$ | 0.8595 | 1.4199 |
| $R^2$ | $1\times10^{-5}$ | 0 | 0 |
| | $2\times10^{-2}$ | 0.0001 | 0.0008 |
| | $1\times10^{-1}$ | 0.0048 | 0.0081 |
| | $2\times10^{-1}$ | 0.0124 | 0.0275 |
| Residual $RMS_{rel}$ | $1\times10^{-5}$ | 0 | 0 |
| | $2\times10^{-2}$ | 0.0089 | 0.0322 |
| | $1\times10^{-1}$ | 0.5290 | 0.9628 |
| | $2\times10^{-1}$ | 0.9788 | 1.8619 |

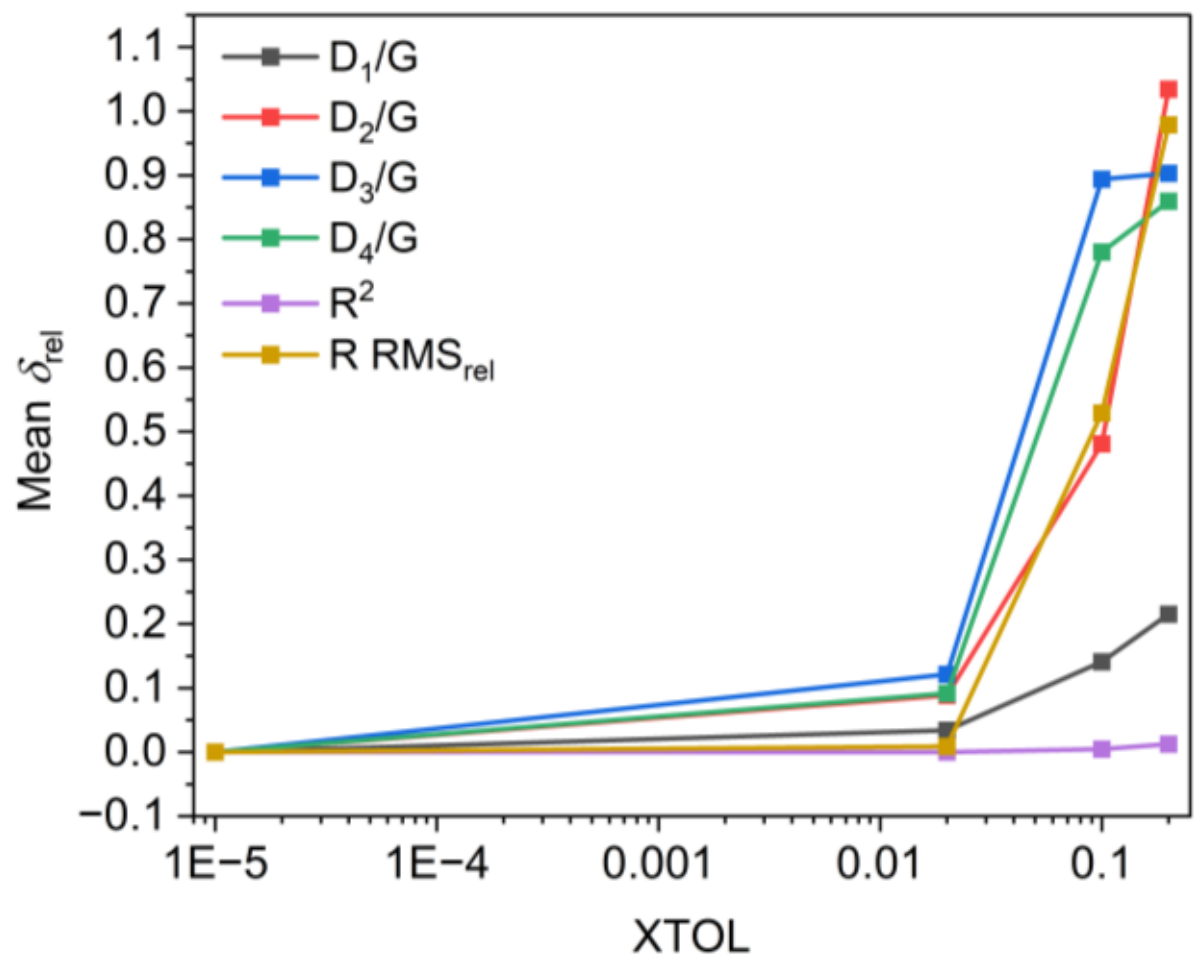

Fig. S5: Average of the sensitivity parameter $\delta_{rel}$ for each Raman subpeak ratios, $R^2$, and residual $RMS_{rel}$ at different levels of deconvolution fitting accuracy.

**Primary particle size distributions**

Figure S6 shows the primary particle size distribution derived from TEM measurements for different plateau temperatures.

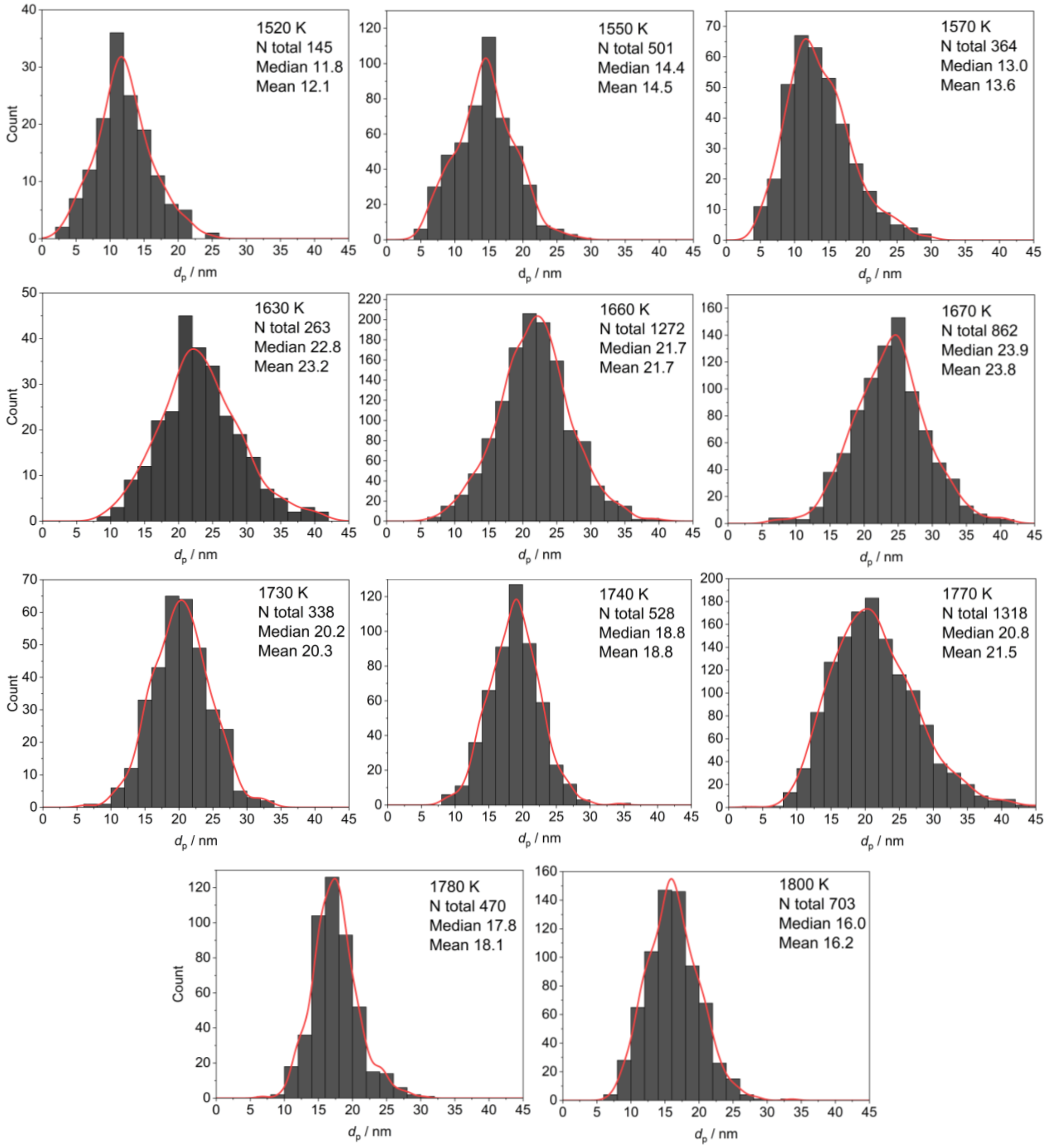

Fig. S6: Primary particle size distributions from TEM measurements of samples generated at different plateau temperatures.